\documentclass[aps,pra,superscriptaddress,amsmath,amssymb,preprintnumbers,floatfix,showpacs,12pt]{revtex4}
\usepackage{amssymb}
\usepackage{epsfig}
\usepackage{subfigure}
\usepackage{graphicx}
\begin{document}
\title{ Quantum statistics and dynamics of nonlinear couplers with
nonlinear exchange }

\author{M. Sebawe Abdalla}

\affiliation{Mathematics Department, College of Science, King Saud
University, P.O. Box 2455, Riyadh 11451, Saudi Arabia}

\author{Faisal A. A. El-Orany\footnote{Permanent address: Suez Canal university, Faculty of
Science, Department of mathematics and computer science, Ismailia,
Egypt.}, J. Pe\v{r}ina  } \affiliation{ Joint Laboratory of Optics
of Palack\'y University and Physical Institute of Academy of
Sciences of Czech Republic, 17.~listopadu~50, 772 07~Olomouc,
Czech Republic. }

\begin{abstract}
In this paper we derive the quantum statistical and dynamical
properties of nonlinear optical couplers composed of two nonlinear
waveguides operating by the second subharmonic generation, which
are coupled linearly through evanescent waves and nonlinearly
through nondegenerate optical parametric interaction. Main
attention is paid to generation and transmission of nonclassical
light, based on a discussion of squeezing phenomenon, normalized
second-order correlation function, and quasiprobability
distribution functions. Initially coherent, number and thermal
states of optical beams are considered. In particular, results are
discussed in dependence on the strength of the nonlinear coupling
relatively to the linear coupling. We show that if the Fock state
$|1\rangle$ enters the first waveguide and the vacuum state
$|0\rangle$ enters the second waveguide, the coupler can serve as
a generator of squeezed vacuum state gevorned by the coupler
parameters. Further, if thermal fields enter initially the
waveguides the coupler plays  similar role as
 a microwave Josephson-junction parametric
amplifier to generate squeezed thermal light.

\end{abstract}
 \maketitle


\section{ Introduction}

In quantum optics many simple quantum systems have been examined
from the point of view of completely quantum statistical
description including not only amplitude and intensity (energy)
development of such systems, but also higher-order moments and
complete statistical behaviour. Such results  have fundamental
physical meaning for interpretation of quantum theory \cite{[1]}
and they are useful for applications in optoelectronics and
photonics as well. These results can be successfully transferred
to more complicated and more practical systems, such as optical
couplers composed of two or more waveguides connected linearly by
means of evanescent waves. The waveguides used can be linear or
nonlinear employing various nonlinear optical processes, such as
optical parametric processes, Kerr effect, Raman or Brillouin
scattering, etc. Such devices play important role in optics,
optoelectronics and photonics as switching and memory elements for
all-optical devices (optical processors and computers). When one
linear and the other nonlinear waveguides are employed, we have a
nonlinear optical coupler producing nonclassical light in the
nonlinear waveguide which can be controlled from the linear
waveguide, i.e. one can control light by light. The generation and
transmission of nonclassical light exhibiting squeezed vacuum
fluctuations and/or sub-Poissonian photon statistics in nonlinear
optical couplers can further be supported when all the waveguides
are nonlinear. The possibility to generate and to transmit
effectively nonclassical light in this way is interesting
especially in optical communication and high-precision
measurements where the reduction of quantum noise increases the
precision. In the present paper nonlinear couplers have been
examined composed of linear and nonlinear waveguides \cite{[2]}
(and references therein) with a particular attention to quantum
statistical properties of such devices \cite{[3],[4],[5],[6],[7]}
related to quantum noise properties. These devices are useful for
generation and transmission of nonclassical light and new
interesting effects can be obtained if phase mismatches are
involved \cite{[8],[9],[10]}. Also Schr\"{o}dinger-cat states can
be transmitted through nonlinear couplers \cite{[11]} and
stability analysis of such devices can be performed \cite{[12]}.

Nonlinear codirectional and contradirectional couplers composed of
two nonlinear waveguides operating by second harmonic generation
or by nondegenerate optical parametric processes can exhibit
interesting switching properties \cite{[13],[14]}.
Quantum-consistent description of contrapropagating beams can be
developed, which permits to formulate the problem in the Hamilton
formalism \cite{[15]}. Phase mismatches inside the nonlinear
waveguides and between them can be taken into account \cite{[10]}.
Interesting results can be obtained for the quantum statistical
properties of nonlinear optical couplers operating by means of
Raman and Brillouin scattering \cite{[7]}.

In this paper we continue in investigation of quantum statistical properties
of nonlinear couplers composed of two waveguides operating by the second
subharmonic generation assuming strong coherent pumping and linear
exchange of energy between waveguides by means of evanescent waves,
however we
additionally take into account the influence of nonlinear coupling of the
parametric type of both the waveguides. In section  2 we describe
dynamics of the system under discussion together with the solution
of the equations of motion, in section  3 we derive squeezing
characteristics of generated light, section  4 is devoted to a
discussion of sub-Poissonian statistics, section  5 includes results for
quasidistribution functions and finally we summarize main conclusions in
section  6.

\section{ Model description and exact solution}

Let us consider a system described by the Hamiltonian $\hat{H}$ such that,

$ {\displaystyle
\frac{\hat{H}}{\hbar} =
\sum_{j=1}^{2}\left\{\omega_{j}\hat{a}_{j}^{\dagger}\hat{a}
_{j}+ \lambda_{j}\left[\hat{a}_{j}^{\dagger 2}\exp (i\mu_{j}t) + {\rm
h.c.}\right]\right\} }
\hfill $

$ {\displaystyle \qquad +\lambda_{3}\left\{\hat{a}_{1}\hat{a}_{2}^{\dagger}
\exp [i\phi_{1}(t)] + {\rm h.c.}\right\} +\lambda_{4}\left\{\hat{a}_{1}
\hat{a_{2}}\exp [-i\phi_{2}(t)] + {\rm h.c.}\right\}, }\hfill (1) $

\noindent where
$\hat{a}_{1}\quad(\hat{a}_{1}^{\dagger})$, $\hat{a}_{2}\quad(\hat{a}
_{2}^{\dagger} $) are annihilation (creation) operators of the fundamental
modes in the first and second waveguides having frequency $\omega_{1}$
and $\omega_{2}$, respectively, $\mu_{j}$ are related with the frequency
of the second-harmonic modes
 described classically as strong coherent
fields, $\phi_{j}(t)$, $j=1,2$, are related to the difference- and
sum-frequencies of modes $1$ and $2$, respectively, $\lambda_{1}$ and
$\lambda_{2}$ are nonlinear coupling constants for the second subharmonic
generation in the first and second waveguides, respectively, $\lambda_{3}$
is the coupling constant for linear exchange between waveguides through
evanescent waves, $\lambda_{4}$ is the coupling constant for the nonlinear
exchange through simultaneous annihilation or creation of a photon in
both the subharmonic modes on expense of pumping and {\rm h.c.} means the
Hermitian conjugate terms (for further details concerning the optical
parametric processes, see \cite{[16]} (Chap. 10)).
When $\mu_{j}=0$ and only the degenerate term
is considered, we have the well-known
Hamiltonian, in the interaction picture,
for squeezed light generation \cite{yun1}, where $\lambda_{1}$ (or
$\lambda_{2}$) represents the coupling constant proportional
to the quadratic susceptibility, of the second-order
nonlinear process (degenerate
parametric down-conversion with classical coherent pumping),
or the coupling constant proportional to the cubic susceptibility,
of the third-order nonlinear process
  (degenerate four-wave mixing with classical coherent pumping) \cite{yam}.
  If additionally $\phi_{1} (t)=\phi_{2}(t)=0$, the Hamiltonian (1)
  represents a mixture of second subharmonic generation, frequency conversion
  and parametric amplification in the interaction picture
\cite{moll1,mish,martin}.

It is important to mention that we
 treat the problem of propagation in the Hamiltonian formalism
neglecting dispersion. Thus if
case
all waves are propagating with the same velocity, time $t$ and space
$z$
relate
by the velocity of propagation $v$, $z=vt$.
Schematically, this
Hamiltonian is represented in Fig. 1.

\begin{picture}(120,70)(-100,6)
\put (20,5){\line(1,0){100}}
\put (20,10){\vector(1,0){80} $\chi^{(2)}$}
\put (20,15){\line(1,0){100}}
\put (20,5){\line(-1,1){35}}
\put (20,15){\line(-1,1){30}}
\put (120,5){\line(1,1){35}}
\put (120,15){\line(1,1){30}}
\put (135,25){\vector(1,1){30}}
\put (180,50){\makebox(0,0){$\hat{a}_{2}(\frac{z}{v})$}}
\put (-15,45){\vector(1,-1){30}}
\put (-27,47){\makebox(0,0){$\hat{a}_{2}(0)$}}
\put (20,-15){\line(1,0){100}}
\put (20,-25){\line(1,0){100}}

\put (20,-65){\line(1,0){100}}
\put (20,-65){\vector(0,1){40}}
\put (120,-65){\vector(0,1){40}}

\put (70,-70){\makebox(0,0){$z=vt$}}
\put (20,-20){\vector(1,0){80}$\chi^{(2)}$}
\put (25,-5){\vector(0,-1){15}$\lambda_{3}$}
\put (25,-5){\vector(0,1){15}}
\put (70,-5){\vector(0,1){15}$\lambda_{4}$}
\put (70,-5){\vector(0,-1){15}}

\put (20,-25){\line(-1,-1){30}}
\put (20,-15){\line(-1,-1){35}}
\put (120,-25){\line(1,-1){30}}
\put (120,-15){\line(1,-1){35}}
\put (175,-55){\makebox(0,0){$\hat{a}_{1}(\frac{z}{v})$}}
\put (135,-35){\vector(1,-1){30}}
\put (-15,-55){\vector(1,1){30}}
\put (-28,-52){\makebox(0,0){$\hat{a}_{1}(0)$}}
\end{picture}
\vspace{1.5in}

{\it Fig.1  Scheme of quantum nonlinear
coupler with linear and nonlinear coupling formed from two nonlinear
waveguides described by the quadratic susceptibility $\chi^{(2)}$.
The beams are described by the photon
annihilation operators as indecated; $z=vt$ is the interaction length.
Both the waveguides are pumped by strong classical coherent waves.
Outgoing fields are examined as single or compound modes by means
of homodyne detection to observe squeezing of vacuum fluctuations,
or by means a set of photodetectors to measure photon antibunchibng
and sub-Poissonian photon statistics in the standard ways.}

 In fact the Hamiltonian
(1) can be regarded as a generalization of the models given in
refs. \cite{moll1,mish,martin,[17],[18],[19]}. For example, if we
take both $\lambda_{1}$ and $\lambda_{2}$ to be zeros, then we
shall be left with the Hamiltonian which describes the back-action
evading amplifiers, where the Hamiltonian in this case can be
constructed by combining parametric amplifiers and parametric
frequency converters with two different coupling parameters. On
the other hand, if we take $\mu_{j}=0$ and drop the time dependent
phases, then the Hamiltonian (1) will be consistent with the
Hamiltonian given in ref. \cite{[20]}, where the wave functions
for both the number state and coherent state and the Green's
function have been obtained. It is also interesting to point out
that the Hamiltonian (1) contains ten generators based on the
group $sp(4,R)$, which represents the most general type of the
two-mode quadratic Hamiltonian \cite{[21]}. This will enable us to
reconsider the problem from Lie algebra point of view, where the
most general solution for the wave functions may be obtained. For
more details, see for example refs. \cite{[22],[23],[24]}, where
the wave function for some special cases of the above Hamiltonian
has been obtained using the Lie algebra technique.

Annihilation and creation operators satisfy the boson commutation relations

$ \displaystyle
\left[\hat{a_{i}},\hat{a_{j}}^{\dagger}\right]=\delta_{ij},
\hfill (2) $

\noindent where $\delta_{ij}$ is the Kronecker delta.

The equations of the motion in the Heisenberg picture for the Hamiltonian
(1) are

$ {\displaystyle
\frac{d\hat{a}_{1}}{dt}=-i\omega_{1}\hat{a}_{1}-2i\lambda_{1} \hat{a}
_{1}^{\dagger} \exp (it\mu_{1}) -i\lambda_{3}\hat{a}_{2}\exp
[-i\phi_{1}(t)]
-i\lambda_{4}\hat{a}_{2}^{\dagger}\exp [i\phi_{2}(t)], } \hfill (3a) $

$ {\displaystyle
\frac{d\hat{a}_{2}}{dt}=-i\omega_{2}\hat{a}_{2}-2i\lambda_{2} \hat{a}
_{2}^{\dagger} \exp (it\mu_{2}) - i\lambda_{3}\hat{a}_{1}\exp
[i\phi_{1}(t)]
-i\lambda_{4}\hat{a}_{1}^{\dagger}\exp [i\phi_{2}(t)]. } \hfill (3b) $

\noindent
 Substituting $\hat{a}_{1}=\hat{A}\exp(\frac{it}{2}\mu_{1})$
and $\hat{a}_{2}= \hat{B}\exp(\frac{it}{2}\mu_{2})$, slowly varying
forms of the operators, having the operators
$\hat{a}_{j}$ as well as $\hat{A}$ and $\hat{B}$ time dependent, equations
 (3) take the form

$ {\displaystyle
\frac{d\hat{A}}{dt}= -i(\omega_{1}+\frac{\mu_{1}}{2})\hat{A} -2i\lambda_{1}
\hat{A}^{\dagger} -i\lambda_{3}\hat{B} \exp \left[ i\frac{(\mu_{2}-\mu_{1})t
}{2}-i\phi_{1}(t)\right] } \hfill $

$ {\displaystyle \qquad -i\lambda_{4}\hat{B}^{\dagger} \exp \left[-i\frac{
(\mu_{1}+\mu_{2})t}{2}+i\phi_{2}(t)\right], }\hfill (4a) $

$ {\displaystyle
\frac{d\hat{B}}{dt} = -i(\omega_{2}+\frac{\mu_{2}}{2})\hat{B}
-2i\lambda_{2}
\hat{B}^{\dagger} -i\lambda_{3}\hat{A} \exp \left[ i\frac{
(\mu_{1}-\mu_{2})t
}{2}+i\phi_{1}(t)\right] } \hfill $

${\displaystyle \qquad -i\lambda_{4}\hat{A}^{\dagger} \exp \left[
-i\frac{
(\mu_{1}+\mu_{2})t}{2 }+i\phi_{2}(t)\right].}$ \hfill (4b)

\noindent Equations (4) with their Hermitian conjugates
give a close system of
four differential equations with time-dependent coefficients  which
cannot be solved directly and hence some restrictions should be considered, so
that we shall consider
 $\phi_{1}(t)=\frac{1}{2}(\mu_{2}-\mu_{1})t$ and
$\phi_{2}(t)=\frac{
1 }{2}(\mu_{2}+\mu_{1})t$.  Then the solutions of the system (4), which
yield the relations between input and output modes, can be
obtained, after some tedious calculations, as

$\hfill \hat{a}_{1}(t)\exp(-it\frac{\mu_{1}}{2})
=\hat{a}_{1}(0)K_{1}(t) +\hat{a}_{1}^{
\dagger}(0)L_{1}(t) +\hat{a}_{2}(0)M_{1}(t)+ \hat{a}_{2}^{
\dagger}(0)N_{1}(t), \hfill (5a) $

$\hfill \hat{a}_{2}(t)
\exp(-it\frac{\mu_{2}}{2})
=\hat{a}_{2}(0)K_{2}(t) +\hat{a}_{2}^{
\dagger}(0)L_{2}(t) +\hat{a}_{1}(0)M_{2}(t)+ \hat{a}_{1}^{
\dagger}(0)N_{2}(t), \hfill (5b) $

\noindent where the time dependent coefficients, which contain all
the features of the structure, are given by

$ {\displaystyle
K_{1}(t)=F_{1}(t)-\frac{i}{2}
\left[ [ k_{+} +k_{-}] G_{1}(t)+[\lambda_{+}\frac{g_{2}}{g_{1}}
+\lambda_{-}]S(t)\right],} \hfill (6a) $

$ {\displaystyle \hfill
L_{1}(t)=-\frac{i}{2}
\left[ [ k_{+} -k_{-}] G_{1}(t)+[\lambda_{+}\frac{g_{2}}{g_{1}}
-\lambda_{-}]S(t)\right], }\hfill (6b) $

$ {\displaystyle  M_{1}(t) =
\frac{1}{2}\left\{\left(1 +
\frac{ g_{2}}{g_{1}}\right)C(t)-
i\left[ [\lambda_{+}+\lambda_{-}]G_{1}(t)+[J_{+}\frac{
g_{2}}{g_{1}} +J_{-}]S(t)\right]\right\},}\hfill (6c) $

$ {\displaystyle
\hfill N_{1}(t) = \frac{1}{2}\left\{\left(1
-\frac{ g_{2}}
{g_{1}}\right) C(t)-i\left[ [\lambda_{+}-\lambda_{-}]G_{1}(t)
+[ J_{+}\frac{g_{2}}{g_{1}}
-J_{-}] S(t)\right]\right\} }, \hfill (6d) $

\noindent whereas

$ {\displaystyle
\hfill
K_{2}(t)=F_{2}(t)-\frac{i}{2}\left[ [J_{+}
+J_{-}] G_{2}(t)+[\lambda_{+}\frac{g_{2}}{g_{1}}
+\lambda_{-}]S(t)\right],}
\hfill (7a) $

$ {\displaystyle
\hfill L_{2}(t)=-\frac{i}{2}\left[ [J_{+}
-J_{-}]G_{2}(t)+[\lambda_{+}-\lambda_{- }\frac{g_{2}}{g_{1}}]S(t)\right],
}\hfill
(7b) $

$ {\displaystyle  \hfill
M_{2}(t)=\frac{1}{2}%
\left\{ \left(1 +\frac{g_{2} }{g_{1}}\right)
C(t)-i\left[ [\lambda_{+}+\lambda_{-}]G_{2}(t)+[k_{+}+k_{-}\frac{g_{2}}
{g_{1}}]S(t)\right]\right\}, } \hfill (7c) $

$ {\displaystyle \hfill N_{2}(t) =
\frac{1}{2}
\left\{ \left( \frac{g_{2}}{g_{1}}-1\right) C(t)-i\left[ [\lambda_{+}-
\lambda_{-}]G_{2}(t)
+[k_{+}-k_{-}\frac{g_{2}}{g_{1}}]S(t)\right]\right\}. } \hfill (7d) $

\noindent In the above equations we have defined

$ {\displaystyle  \lambda_{\pm}=\lambda_{3}\pm \lambda_{4}, }\hfill $

$ {\displaystyle k_{\pm}=
\omega_{1}+\frac{1}{2}\mu_{1}\pm2\lambda_{1}, }
\hfill $

$ {\displaystyle J_{\pm}=\omega_{2}+\frac{1}{2}\mu_{2}\pm2\lambda_{2}, }
\hfill $

$ {\displaystyle  g_{1}=k_{-}\lambda_{+}+\lambda_{-}J_{+}, }\hfill $

$ {\displaystyle  g_{2}=k_{+}\lambda_{-}+\lambda_{+}J_{-}, }\hfill (8) $

\noindent and

$ {\displaystyle  F_{1}(t)= \cos (t\bar{\Omega}_{1}) \cos ^{2}\theta +
\cos(t\bar{\Omega}_{2})\sin ^{2}\theta ,} \hfill (9a) $

$ {\displaystyle  F_{2}(t)=\cos (t\bar{\Omega}_{2}) \cos ^{2}\theta
+\cos (t
\bar{\Omega}_{1}) \sin ^{2}\theta ,} \hfill (9b) $

$ {\displaystyle  G_{1}(t)=\frac{\sin
(t\bar{\Omega}_{1})}{\bar{\Omega}_{1} }
\cos ^{2}\theta + \frac{\sin (t\bar{\Omega}_{2})}{\bar{\Omega}_{2}} \sin
^{2}\theta , } \hfill (9c) $

$ {\displaystyle \hfill G_{2}(t)=\frac{\sin
(t\bar{\Omega}_{2})}{\bar{\Omega}
_{2} }\cos ^{2}\theta + \frac{\sin
(t\bar{\Omega}_{1})}{\bar{\Omega}_{1}}
\sin ^{2}\theta , }\hfill (9d) $

$ {\displaystyle \hfill
C(t)=\frac{1}{2}\sqrt{\frac{g_{1}}{g_{2}}}\left[\cos
(t\bar{\Omega}_{2}) -\cos (t\bar{\Omega}_{1})\right]\sin (2\theta ) , }
\hfill (9e) $

$ {\displaystyle  \hfill
S(t)=\frac{1}{2}\sqrt{\frac{g_{1}}{g_{2}}}\left[
\frac{\sin (t\bar{\Omega}_{2})}{\bar{\Omega}_{2} } - \frac{\sin
(t\bar{\Omega
}_{1})}{\bar{\Omega}_{1} } \right]\sin (2\theta ), } \hfill (9f) $

\noindent where we have introduced the abbreviations

$ {\displaystyle \theta=\frac{1}{2}\tan
^{-1}\left(\frac{2\sqrt{g_{1}g_{2}}
} {J_{-}J_{+}-k_{-}k_{+}}\right) }, \hfill (10a) $

$\bar{\Omega}_{1}=\left[ \Omega_{1}^{2}\cos^{2}\theta +
\Omega_{2}^{2}\sin^{2}\theta -\sqrt{g_{1}g_{2}}\sin
(2\theta)\right]^{\frac{
1 }{2}}, \hfill (10b) \label{24} $

$\bar{\Omega}_{2}=\left[ \Omega_{2}^{2}\cos^{2}\theta +
\Omega_{1}^{2}\sin^{2}\theta +\sqrt{g_{1}g_{2}}\sin
(2\theta)\right]^{\frac{
1 }{2}}, \hfill (10c) \label{25} $

\noindent with $\Omega_{1}^{2}=\lambda_{-}\lambda_{+}+k_{-}k_{+}$ and $
\Omega_{2}^{2}=\lambda_{-}\lambda_{+}+J_{-}J_{+}$.
\setcounter{equation}{10}

One can see from this solution that when $\bar{\Omega}_{1}$ and
$\bar{\Omega}_{2}$ are real, the coupler switches the energy between the
modes which  propagate inside since the solution will include trigonometric
functions \cite{moll1}. Nevertheless, if $\bar{\Omega}_{1}$ and
$\bar{\Omega}_{2}$ are pure imaginary, the Heisenberg solutions
attribute hyperbolic functions, which are growing rapidly with time, and the
coupler operates as amplifier for the input modes \cite{tuck1}. So that
 the behaviour of the coupler will be indicated essentially by the
relation between coupling constants.

 For the time dependent coefficients, we can easily obtain the
following relations

$ {\displaystyle
|K_{j}(t)|^{2}+|M_{j}(t)|^{2}=1+|L_{j}(t)|^{2}+|N_{j}(t)|^{2};
\quad j=1,2, }
\hfill (11a) $

$ {\displaystyle
K_{1}(t)N_{2}(t)+M_{1}(t)L_{2}(t)=N_{1}(t)K_{2}(t)+L_{1}(t)M_{2}(t),
}\hfill
(11b) $

$ {\displaystyle
K_{1}(t)M_{2}^{*}(t)+M_{1}(t)K_{2}^{*}(t)=L_{1}(t)N_{2}^{*}(t)
+N_{1}(t)L_{2}^{*}(t) }, \hfill (11c) $

\noindent in correspondence to boson commutation rules (2).

In what follows, we shall employ the results obtained in the present section
to treat the squeezing phenomena, normalized second-order correlation
function, as
well as quasiprobability distribution functions for the model under
consideration.

\section{ Squeezing phenomenon}

Squeezing  is a pure nonclassical phenomenon and
squeezed states have less noise in one field quadrature than a coherent
state. On the other hand, this means that there is an excess of noise in
the
conjugate quadrature, since the product of canonically conjugate
variances
must satisfy the uncertainty relation.
This light has a lot of
 applications, e.g. in optical
communication networks \cite{yu1}, in interferometric techniques
\cite{ca1}, and in optical waveguide tap \cite{sha}. Generation of
squeezed light has been observed in many optical processes
\cite{lu1,pi1}. Investigation of the squeezing properties of the
radiation field is a central topic in quantum optics which can be
measured by homodyne detection where the signal is superimposed on
a strong coherent beam of the local oscillator.

For this purpose we define the position and momentum operators, which are
related to the conjugate electric and magnetic field operators
$\hat{E}
$ and $\hat{H}$ of electromagnetic field,
for each mode
 in
terms of $\hat{a}_{j}(t)$ and $\hat{a}_{j}^{\dagger}(t)$ as

$ {\displaystyle \hat{X}_{j}(t)=\frac{1}{2}\left[ \hat{a}_{j}(t)\exp
(\frac{
it\mu_{j}}{2}) +{\hat{a}_{j}}^{\dagger}(t)\exp (\frac{-it\mu_{j}}{2})
\right],} \hfill (12) $

$ {\displaystyle \hat{Y}_{j}(t)=\frac{1}{2i}\left[ \hat{a}_{j}(t)\exp
(\frac{it\mu_{j} }{2}) -\hat{a_{j}}^{\dagger}(t)\exp (\frac{-it\mu_{j}}{2})
\right],
} \hfill (13) $

\noindent where we have considered $\mu_{j} (t)$ to be the phase of the
local
oscillator, without loss of generality, to cancel the high frequency terms,
and $j=1,2$ stands for mode 1 and mode 2, respectively.
These
operators satisfy the commutation relations

$ {\displaystyle \left[\hat{X}_{j}(t),\hat{Y}_{j}(t)\right]=\frac{i}{2},}
\hfill (14) $

\noindent so that the uncertainty relations are

$ {\displaystyle \triangle \hat{X}_{j}(t)\triangle\hat{Y}_{j}(t)\geq
\frac{1}{4},}
\hfill (15) $

\noindent with $\triangle
\hat{X}_{j}(t)=
\left[ \langle\left(\triangle \hat{X}_{j}(t)\right)^{2}\rangle \right]^
{\frac{1}{2}}
=[\langle\hat{X}^{2}_{j}(t)\rangle-\langle
\hat{X}_{j}(t) \rangle^{2} ]^{\frac{1}{2}}$.

One of the following squeezing conditions for each mode can occur,

${\displaystyle  S_{j}(t)=4\langle\left(\triangle \hat{X}_{j}(t)\right)^{2}\rangle
- 1<0,} $

$Q_{j}(t)=4\langle\left(\triangle \hat{Y}_{j}(t)\right)^{2}\rangle- 1<0,
\hfill (16) $

\noindent i.e. negative values of these quantities express squeezing of
vacuum fluctuations.
Here we study squeezing phenomenon when the modes are initially prepared
in thermal-states (or in number states since both of these two cases,
number states and thermal-states, have
identical quadrature variances) with the average  thermal  photon
numbers
$\bar{n}_{j},\quad j=1,2$ as well as in the coherent states. More details
on the evolution of thermal light in the model under discussion will be
adopted in section 5.
Now for the quantities $S_{j}(t)$ and $Q_{j}(t)$, provided that
both the modes are initially in the thermal states, we have for the first
mode the following expressions

$ {\displaystyle S_{1}(t) = 2\bar{n}_{1}[|L_{1}(t)|^{2}+|K_{1}(t)|^{2}]
+2\bar{n}_{2}[|N_{1}(t)|^{2}+|M_{1}(t)|^{2}]
+2|L_{1}(t)|^{2}+2|N_{1}(t)|^{2}
} \hfill$

$ {\displaystyle
+2(2\bar{n}_{1}+1)[ L_{1}(t)K_{1}(t) + {\rm c.c.}]
+2(2\bar{n}_{2}+1)[ M_{1}(t)N_{1}(t) + {\rm c.c.}],} \hfill
(17) $

$\hfill $

$ {\displaystyle Q_{1}(t) = 2\bar{n}_{1}[|L_{1}(t)|^{2}+|K_{1}(t)|^{2}]
+2\bar{n}_{2}[|N_{1}(t)|^{2}+|M_{1}(t)|^{2}]
+2|L_{1}(t)|^{2}+2|N_{1}(t)|^{2}
} \hfill$

$ {\displaystyle
-2(2\bar{n}_{1}+1)[ L_{1}(t)K_{1}(t) + {\rm c.c.}]
-2(2\bar{n}_{2}+1)[ M_{1}(t)N_{1}(t) + {\rm c.c.}],} \hfill
(18) $

\noindent where c.c. means the complex conjugate terms.  The
corresponding expressions for the second mode can be obtained from (17)
and (18) by using  the interchange $1\leftrightarrow 2$. However,
the other expressions related to the injected coherent light initially
in the coupler are the same (17) and (18) but  just put $\bar{n}_{j}=0$.

It is known that the nonlinear coupler is a source of optical
fields, the statistical properties of which are changed as a
result of the linear and nonlinear interaction inside and between
waveguides. Consequently, one can generate nonclassical light from
one input and, in addation, it can be switched.
\setcounter{figure}{1}
\begin{figure}[h]%
  \centering
  \subfigure[]{\includegraphics[width=8cm]{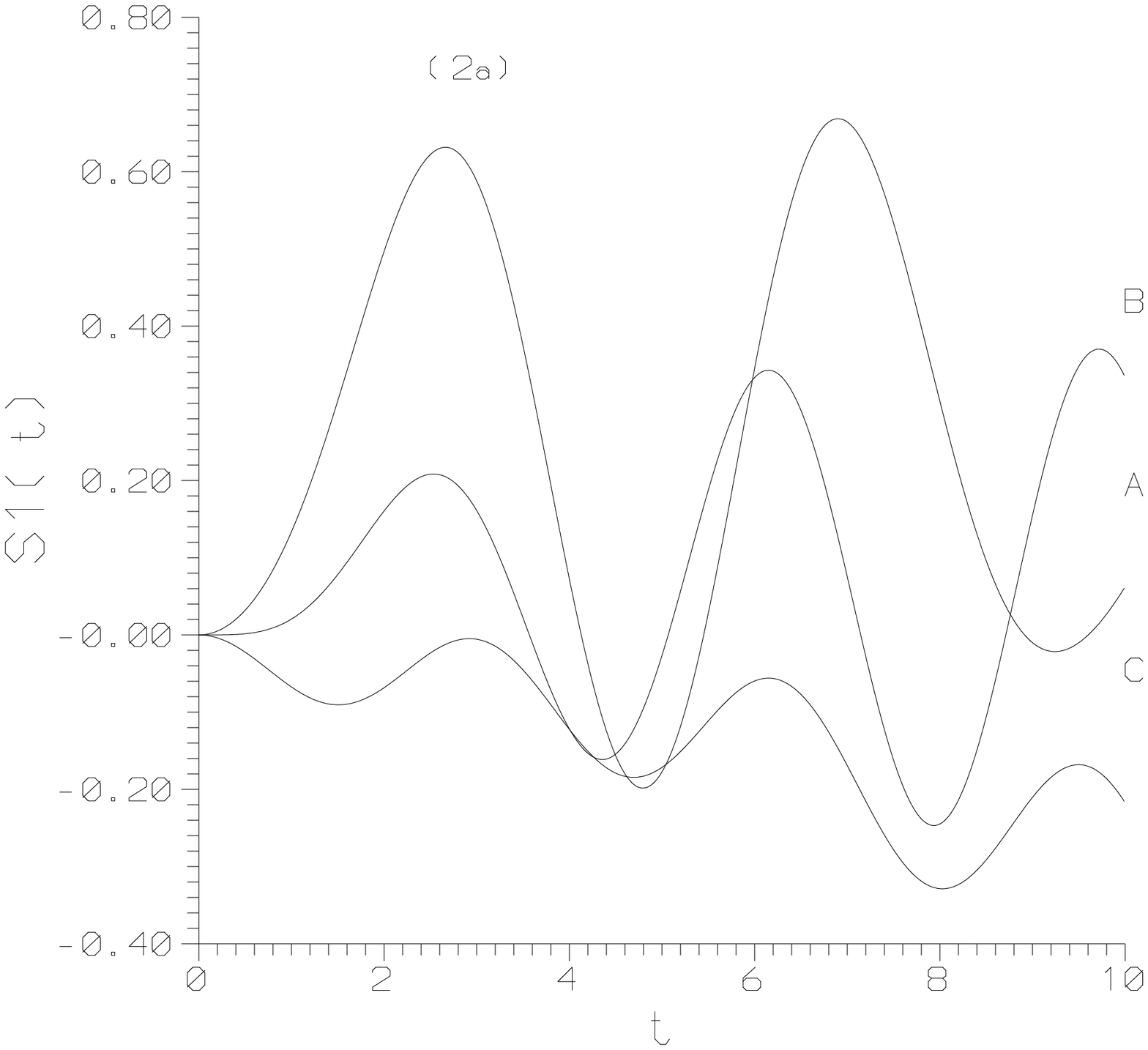}}
 \subfigure[]{\includegraphics[width=8cm]{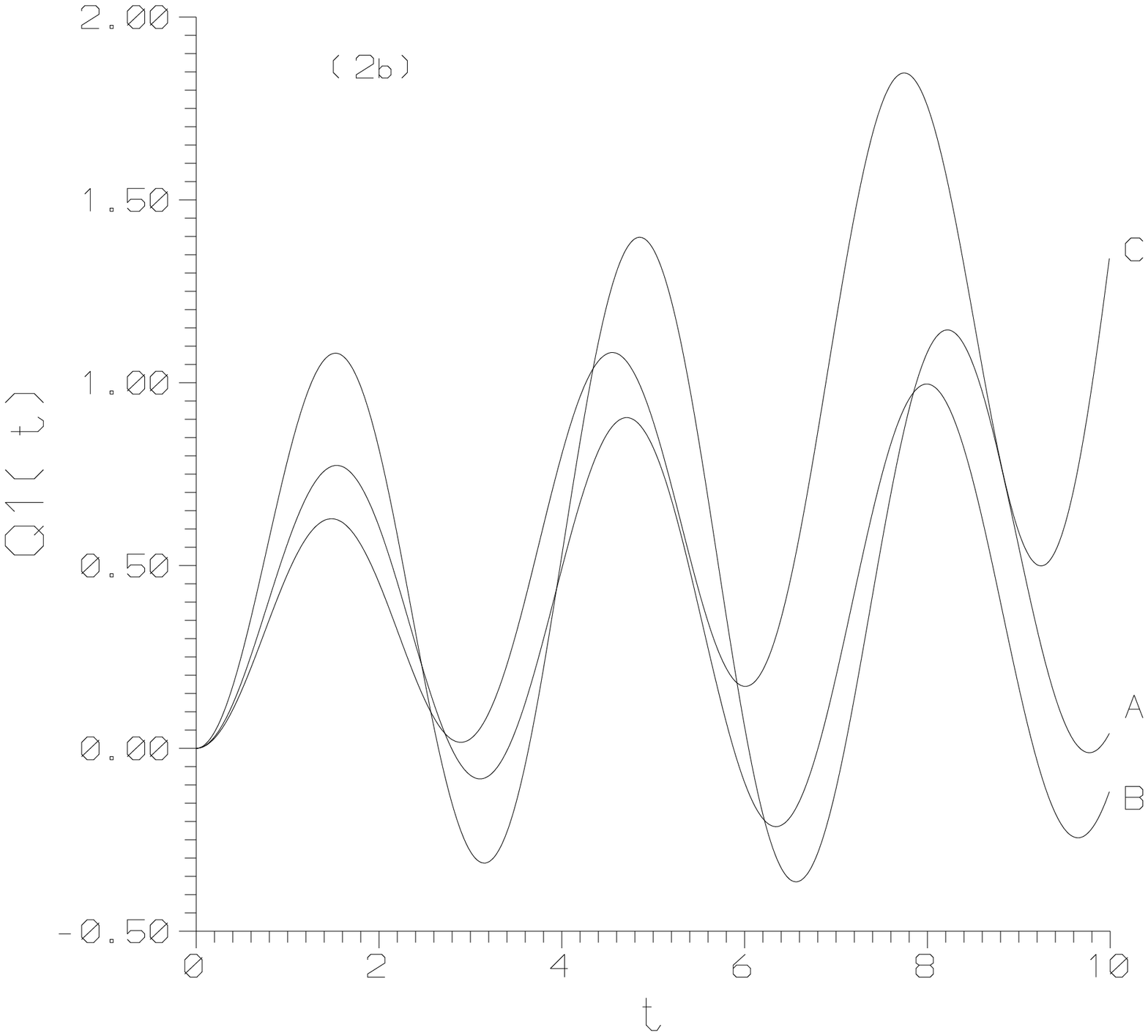}}
 \subfigure[]{\includegraphics[width=8cm]{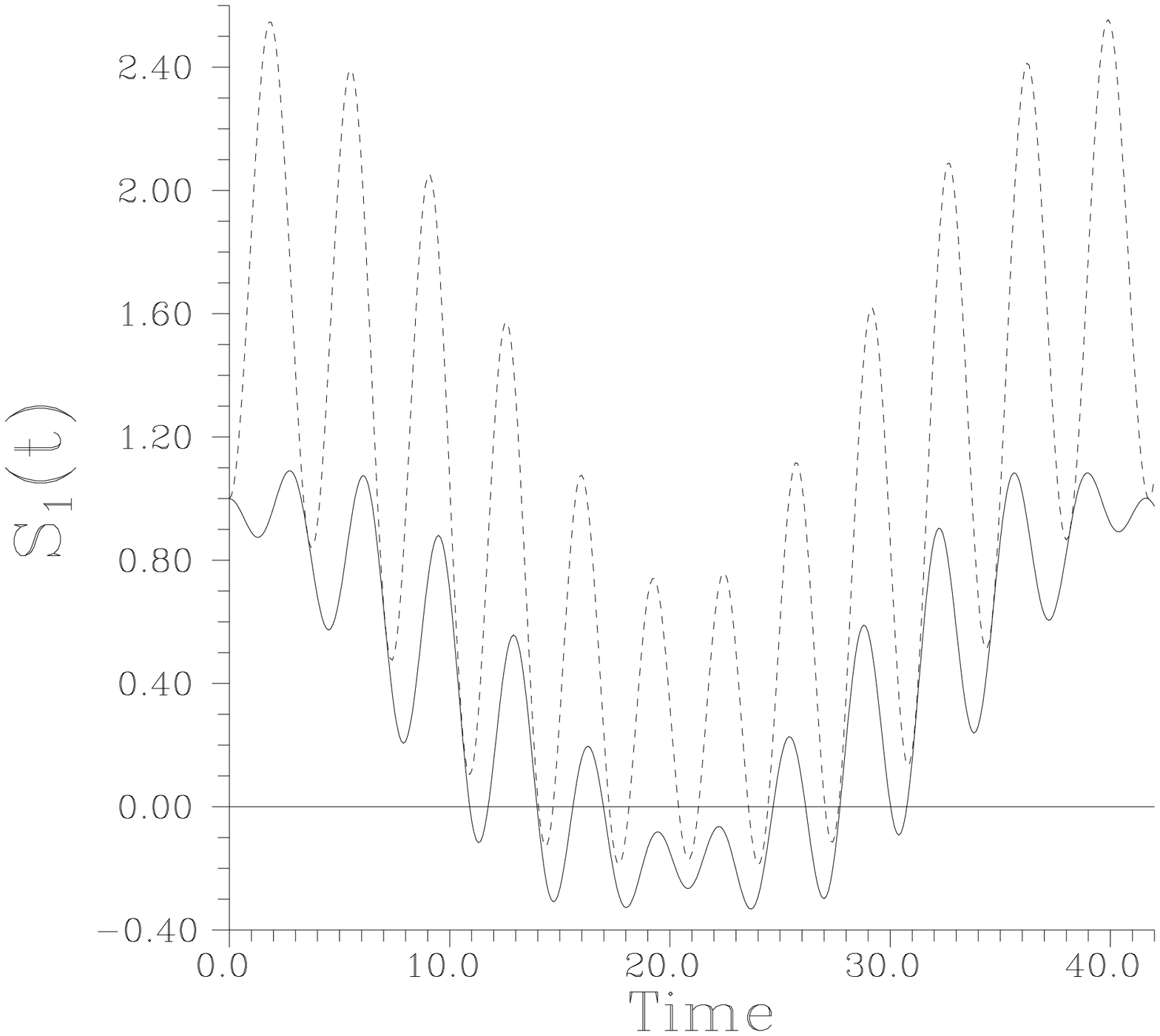}}
    \caption{
 Squeezing phenomenon for mode 1 when the modes are initially
 in coherent light and in thermal light. For initial coherent light
: a) for the first
component $%
S_{1}(t)$; b) for the second component $Q_{1}(t)$; $\lambda_{3}=1$
for all curves; curves A, B and C are corresponding to
$\lambda_{1}= \lambda_{2}=\lambda_{4}=0.25$, $\lambda_{1} =
\lambda_{2}=\lambda_{4}=0.2$ and $\lambda_{1}=0.17$,
$\lambda_{2}=\lambda_{4}=0.2$, respectively. For initial thermal
light: c) the first component $S_{1}(t)$ with $\bar{n}_{1}=0.5$,
$\bar{n}_{2}=0.5$ (solid curve), $1.5$ (dashed curve) and the
coupling constants $\lambda_{j}$ are the same as those for the
curve C when the light is initially coherent; straight line has
been put to show the bound of squeezing.  }
  \label{fig2}
\end{figure}

We have plotted $S_{1}(t), Q_{1}(t)$ in Figs. 2a,b and $S_{2}(t), Q_{2}(t)$
in Figs. 3a,b, when the initial light is coherent, for different values of
$\lambda_{k}$. Further we have chosen $
\lambda_{3}=1$ for all curves and for the curve $A: \lambda_{1}=\lambda_{2}
=\lambda_{4}=0.25$; for the curve $B: \lambda_{1}=\lambda_{2}
=\lambda_{4}=0.20$, and for the curve $C: \lambda_{1}=0.17,\lambda_{2}
=\lambda_{4}=0.2$. On the other hand, Fig. 2c gives
$S_{1}(t)$ (first mode) when the initial light is thermal light with
coupling constants as those for the curve C, where $\bar{n}_{1}=0.5$ and
$\bar{n}_{2}=0.5$ (solid curve), 1.5 (dashed curve); and straight line
 shows the bound of squeezing of the curves.
Firstly, we start our discussion by studying the case of input coherent
light.
\begin{figure}[h]%
  \centering
  \subfigure[]{\includegraphics[width=8cm]{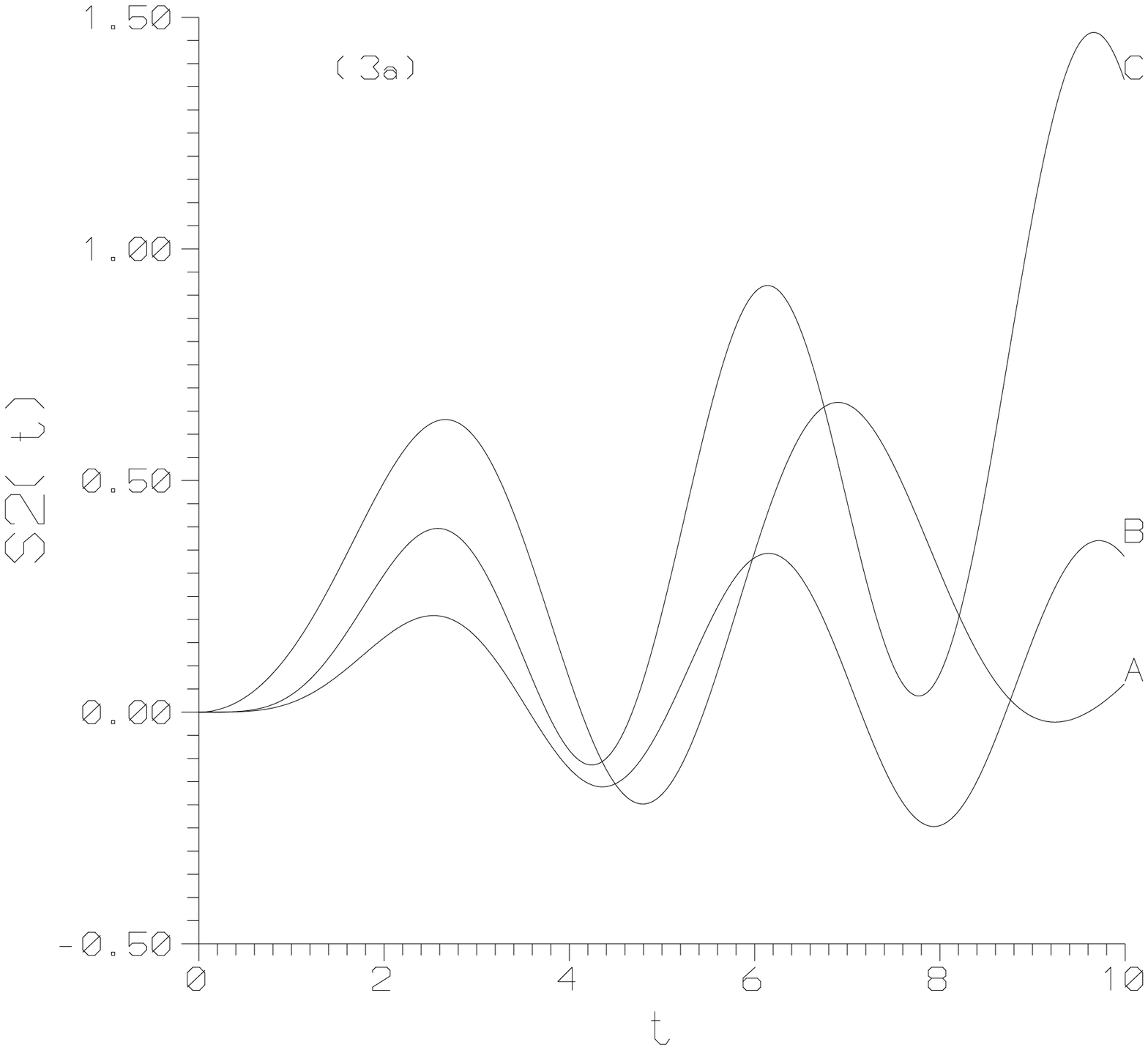}}
 \subfigure[]{\includegraphics[width=8cm]{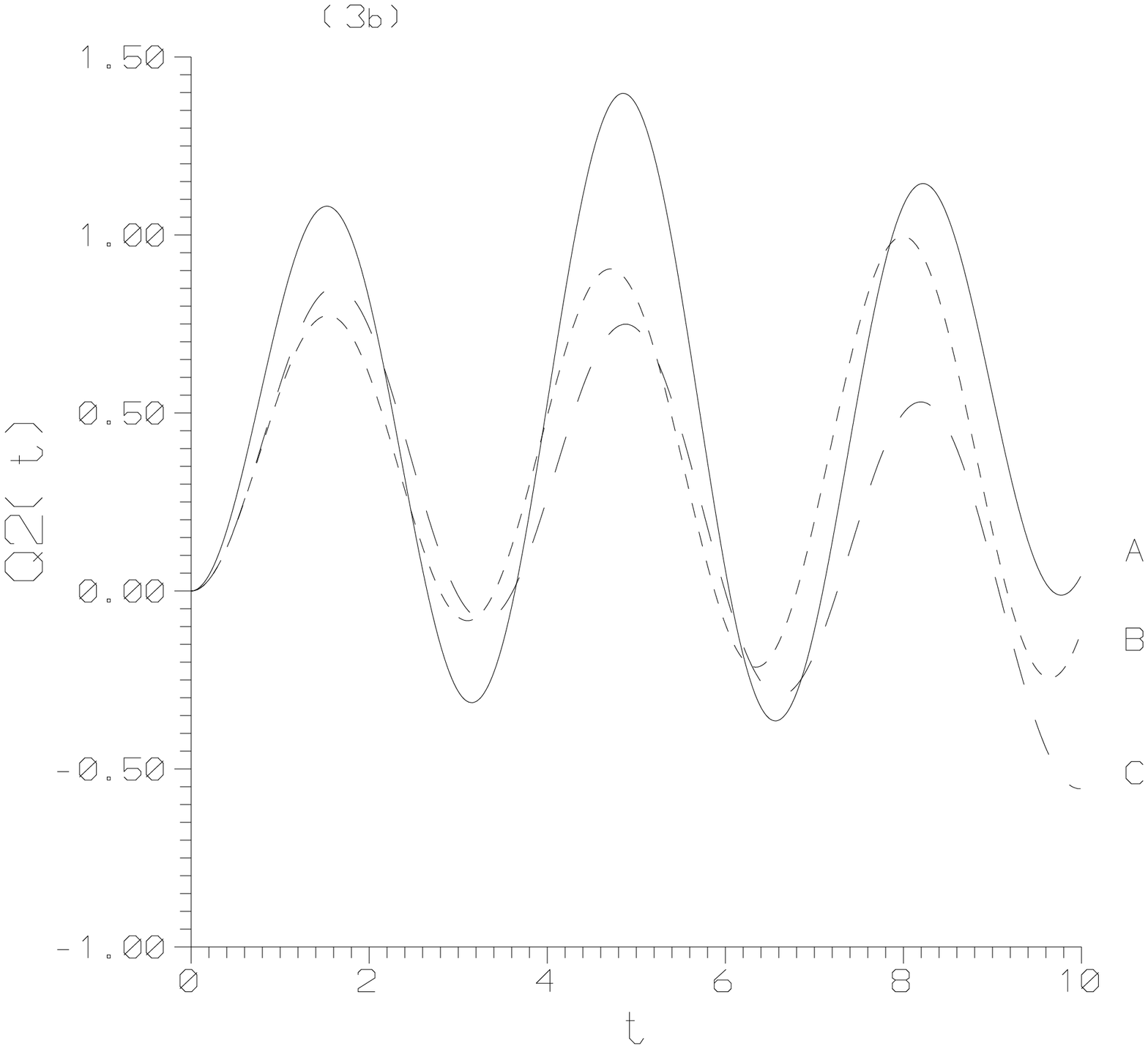}}
     \caption{
Squeezing phenomenon for mode 2: a) for the first component $
S_{2}(t)$; b) for the second component $Q_{2}(t)$; the values of
the parameters $\lambda_{k}$ are as in Fig. 2.  }
  \label{fig3}
\end{figure}
From these figures we can see how the coherent states, which are
minimum-uncertainty states, evolve in the coupler to produce squeezed light.
We can observe  the oscillatory behaviour in these curves, showing
that squeezing can be switched from one waveguide to the other in the
course of time during power transfer. Moreover, squeezing can be
interchanged between the two quadratures of the same waveguide.
More precisely, for mode 1, squeezing can occur for all selected values
of $\lambda_{k}$ in $S_{1}(t)$, but in $Q_{1}(t)$ only curves A, B can
exhibit squeezing, as shown in Figs. 2a,b, which reflects the dependence of
nonclassical behaviour on the strength of subharmonic generation. For mode 2
we can see squeezing in all curves in both the quadratures, as shown in
Figs. 3a,b. It can be easily seen that the amount of squeezing is sensitive
to the strength of coupling $\lambda_{k}$ and that in general its values in
the second component are more pronounced than those in the first one.
Now if we turn our attention to the case of injected thermal light, i.e.
Fig. 2c, we can observe that squeezing is available in the large
interaction time. Further, $S_{1}(t)$ exhibits oscillatory behaviour and
it evolved from unsqueezed values in the short range of interaction time,
 owing to the fact that thermal-states are not minimum-uncertainty states, into
 squeezed values and eventually unsqueezed values can be recovered. Indeed, we noted
 numerically that this behaviour is periodically recovered  with the
time.   Moreover, by comparing
the dashed curve with the solid one, we can see that increasing
of the photon
 number in the second waveguide causes decreasing of the amount of squeezing in
 the first one. This is related with the effect of evanescent waves
 bewteen waveguides and shows how one can  control light by light in the
 coupler. Finally, we can conclude that by controlling  the input
 average thermal
 photon number and  the interaction time (or on the length of the
 coupler), the interaction under consideration can generate squeezed
 thermal light.
It is worthwhile to refere to \cite{kim2},
where more discussions related to  squeezed thermal states
are
given.
Furthermore, squeezing of thermal radiation field
has
been already produced in a microwave Josephson-junction parametric
amplifier
\cite{yur}, where a thermal input field has been introduced to the squeezing
device
and the generated field has exhibited noise reduction.

\section{ Second-order correlation function}

Starting with the experiment of Hanbury Brown and Twiss, strong interest in
the photon-counting statistics of optical fields began. Traditional
diffraction and interference experiments and spectral measurements may be
considered as being performed in the domain of one photon or linear optics.
The theory of higher-order optical phenomena, described by higher-order
correlation functions of the electromagnetic field, was founded by Glauber
\cite{[25]}, who introduced the measure of super-Poissonian statistics
 (classical phenomenon) and
sub-Poissonian statistics (nonclassical phenomenon) of photons in any state,
which is
given by the normalized normal second-order correlation function defined as

$ {\displaystyle g_{j}^{(2)}(t) = \frac{\langle \hat{a}_{j}^{\dagger 2}(t)
\hat{a}_{j}^{2}(t)\rangle } {\langle \hat{a}_{j}^{\dagger}(t) \hat{a}
_{j}(t)\rangle^{2}} } \hfill $

$ {\displaystyle  \qquad =1+\frac{\langle (\triangle\hat{n}
_{j}(t))^{2}\rangle - \langle \hat{a}_{j}^{\dagger}(t)
\hat{a}_{j}(t)\rangle
}{\langle \hat{a}_{j}^{\dagger}(t) \hat{a}_{j}(t)\rangle^{2}}, }
\hfill (19)$

\noindent where the subscript $j$ relates to the $j$th mode and $\langle
(\triangle \hat{n}_{j}(t))^{2}\rangle$ are the photon number variances, which
can be obtained from the relation

$ {\displaystyle \langle (\triangle\hat{n}_{j}(t))^{2}\rangle =
\langle(\hat{a}_{j}^{\dagger}(t) \hat{a}_{j}(t))^{2}\rangle -\langle \hat{a}
_{j}^{\dagger}(t) \hat{a}_{j}(t)\rangle^{2}. } \hfill (20) $

\noindent Then it holds that $g_{j}^{(2)}(t)<1$ for
sub-Poissonian distribution of photons, $g_{j}^{(2)}(t)>1$ for
super-Poissonian distribution of photons and when $g_{j}^{(2)}(t) =1$
Poissonian distribution occurs. The degree of coherence $g^{(2)}_{j}(t)$
can be measured by a set of two detectors.
 An application of radiation exhibiting the
sub-Poissonian
statistics to optical communications has been considered in \cite{[31]}.

The most familiar quantum states from the earlier days of quantum
mechanics are coherent and number states. Following the
development of the quantum theory of radiation and with the advent
of the laser, the coherent states of the field, that mostly
describe a classical electromagnetic field, were widely studied.
These states are minimum-uncertainty states and have Poissonian
distribution of photons and they may be evolved in the nonlinear
optical coupler to generate nonclassical light. On the other hand,
number states are purely nonclassical states (they always exhibit
sub-Poissonian statistics) and there is great interest for their
preparation and quantum non-demolition detection
\cite{[32],[33],[34]}, because they exhibit the maximum channel
capacity, i.e. they provide the maximum of information that can be
transmitted by a single photon, and the minimum time-energy
product in optical communications \cite{[35]}.

Here we shall study the intensities of the fields as well as
 the normalized normal second-order correlation function
for mode 1 when both the modes are initially in the coherent states $
|\alpha\rangle_{1}, |\beta\rangle_{2}$ or in the number states $
|n\rangle_{1},|m\rangle_{2}$. Then the photon number variance in the
coherent state is given by

$ {\displaystyle \langle (\triangle\hat{n}_{j}(t))^2\rangle_{{\rm coh}} =
[V^{(j)2}_{1}(t) +4|V^{(j)}_{4}(t)|^{2}]|\alpha|^{2}
+[V^{(j)2}_{2}(t)+4|V^{(j)}_{5}(t)|^{2}]|\beta|^{2} } \hfill $

$ {\displaystyle \qquad + [|V^{(j)}_{7}(t)|^2
+|V^{(j)}_{6}(t)|^{2}](|\alpha|^{2}+|\beta|^{2})
+[|V^{(j)}_{6}(t)|^{2}+2|V^{(j)}_{4}(t)|^{2}+2|V^{(j)}_{5}(t)|^{2}]}
\hfill $

$ {\displaystyle \qquad +\Bigl\{
\alpha^{2}[2V^{(j)}_{1}(t)V^{*(j)}_{4}(t)
+V^{*(j)}_{7}(t)V^{(j)}_{6}(t)]+
\beta^{2}[2V^{(j)}_{2}(t)V^{*(j)}_{5}(t) +V^{(j)} _{7}(t)V^{(j)}_{6}(t)] }
\hfill $

$ {\displaystyle \qquad +\alpha\beta
[V^{(j)}_{1}(t)V^{(j)}_{6}(t)+2V^{*(j)}_{4}(t)V^{(j)}_{7}(t)
+2V^{*(j)}_{5}(t)V^{*(j)}_{7}(t) +V^{(j)}_{2}(t)V^{(j)}_{6}(t)] } \hfill $

$ {\displaystyle \qquad +\alpha^{*}\beta [V^{(j)}_{1}(t)V^{(j)}_{7}(t)
+2V^{(j)}_{4}(t)V^{(j)}_{6}(t) +2V^{*(j)}_{5}(t)V^{*(j)}_{6}(t)
+V^{(j)}_{2}(t)V^{(j)}_{7}(t)]+ {\rm c.c.}\Bigr\}, }\hfill (21) $

\noindent while the expectation value of the photon number is

$ {\displaystyle \langle \hat{a}_{j}^{\dagger}(t)
\hat{a}_{j}(t)\rangle_{{\rm coh}} = |\alpha|^{2}V^{(j)}_{1}(t)
+|\beta|^{2}V^{(j)}_{2}(t)+V^{(j)}_{3}(t) }\hfill $

$ {\displaystyle \qquad +\left[
\alpha^{2}V^{*(j)}_{4}(t)+\beta^{2}V^{*(j)}_{5}(t)+\alpha^{*}\beta
V^{(j)}_{7}(t) +\alpha\beta V^{(j)}_{6}(t)+{\rm c.c.} \right]. }\hfill
(22)$

\noindent For initial number state we find the photon number variance in
the form

$ {\displaystyle \langle (\triangle\hat{n}_{j}(t))^{2}\rangle_{n} =
2|V^{(j)}_{4}(t)|^{2}(n^{2}+n+1) +2|V^{(j)}_{5}(t)|^{2} (m^{2}+m+1)}
\hfill $

$ {\displaystyle \qquad
+(|V^{(j)}_{6}(t)|^{2}+|V^{(j)}_{7}(t)|^{2})(n+m+2mn), } \hfill (23) $

\noindent while the expectation value of the photon number is

$ {\displaystyle
 \langle \hat{a}_{j}^{\dagger}(t) \hat{a}_{j}(t)\rangle_{n}
 =n V^{(j)}_{1}(t) +mV^{(j)}_{2}(t)+V_{3}^{(j)}(t),} \hfill (24) $

\noindent where

$ {\displaystyle V^{(j)}_{1}(t)=|K_{j}(t)|^{2}+|L_{j}(t)|^{2},} \hfill
(25a) $

$ {\displaystyle V^{(j)}_{2}(t)=|M_{j}(t)|^{2}+|N_{j}(t)|^{2},} \hfill
(25b) $

$ {\displaystyle V^{(j)}_{3}(t)=|N_{j}(t)|^{2}+|L_{j}(t)|^{2},} \hfill
(25c) $

$ {\displaystyle V^{(j)}_{4}(t)=K_{j}^{*}(t)L_{j}(t), } \hfill (25d) $

$ {\displaystyle V^{(j)}_{5}(t)=M_{j}^{*}(t)N_{j}(t), } \hfill (25e) $

$ {\displaystyle V^{(j)}_{6}(t)=K_{j}(t)N_{j}^{*}(t) +L_{j}^{*}(t)M_{j}(t),}
\hfill (25f) $

$ {\displaystyle V^{(j)}_{7}(t)=M_{j}(t)K_{j}^{*}(t)
+N_{j}^{*}(t)L_{j}(t),}
\hfill (25g) $

\noindent and $j=1,2$ corresponding to first and second mode, respectively.
\begin{figure}[h]%
  \centering
  \subfigure[]{\includegraphics[width=8cm]{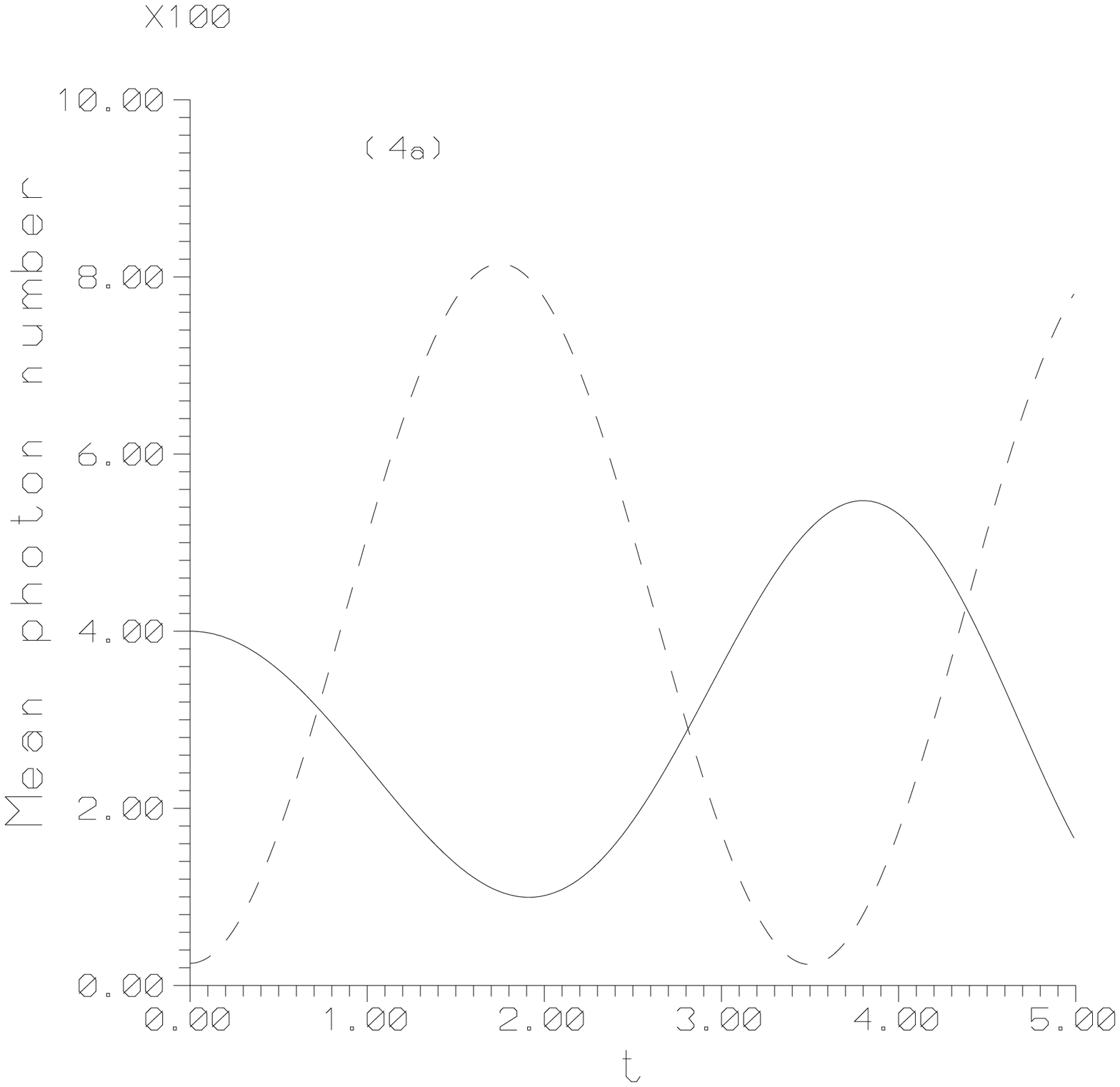}}
 \subfigure[]{\includegraphics[width=8cm]{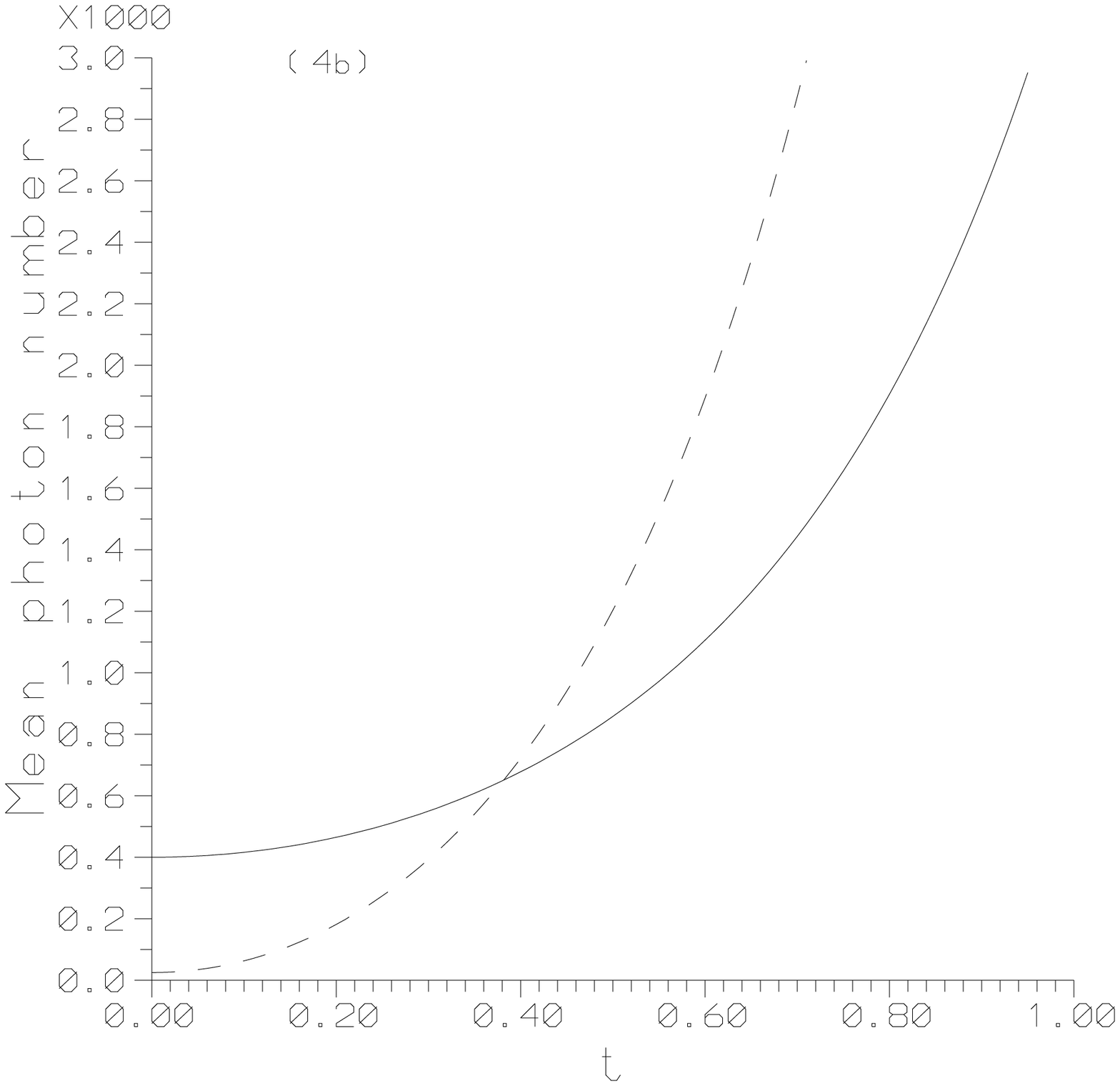}}
     \caption{
 Mean photon number against time $t$ for mode 1 (solid
curve) and mode 2 (dashed curve)  when both the modes are
initially in the coherent states with $\alpha =20,
\beta=5,\lambda_{1}=0.17,\lambda_{2}=0.2,\lambda_{3}=1$: a)
$\lambda_{4}=0.2$; b) $\lambda_{4}=2$.  }
  \label{fig5}
\end{figure}
It is important to study the evolution of the mean photon numbers
(intensities)
$\langle \hat{a}^{\dagger}_{j}(t) \hat{a}_{j}(t)\rangle$ inside the
waveguides of the coupler to visualize how the energy is
exchanged between the waveguides. For this purpose we show Fig. 4
in which
the mean photon number (22) of the beams
is plotted against the time $t$ for
shown values of the parameters. The solid and dashed curves are
related to the
first and second beams, respectively.
 We note that the essential for the behaviour of the coupler
 under consideration is
relation of powers
of the linear ($\lambda_{3}$) and nonlinear ($\lambda_{4}$)
coupling constants.
 To be more specific, for $\lambda_{3}>\lambda_{4}$ (Fig. 4a), the intensities
 evelve oscillatory with time $t$, which means that the
periodic power
 transfer occurs between waveguides and the coupler operates as
an optical
 switcher. Further,  at certain values of time,
corresponding to intersections of the two curves,
all energy in the coupler becomes
 equally shared between the propagating modes.  However, for
 $\lambda_{3}<\lambda_{4}$ (Fig. 4b), the initial
 intensities are amplified in the course of time and the coupler
 operates as an amplifier for input modes.

\begin{figure}[h]%
  \centering
  \subfigure[]{\includegraphics[width=8cm]{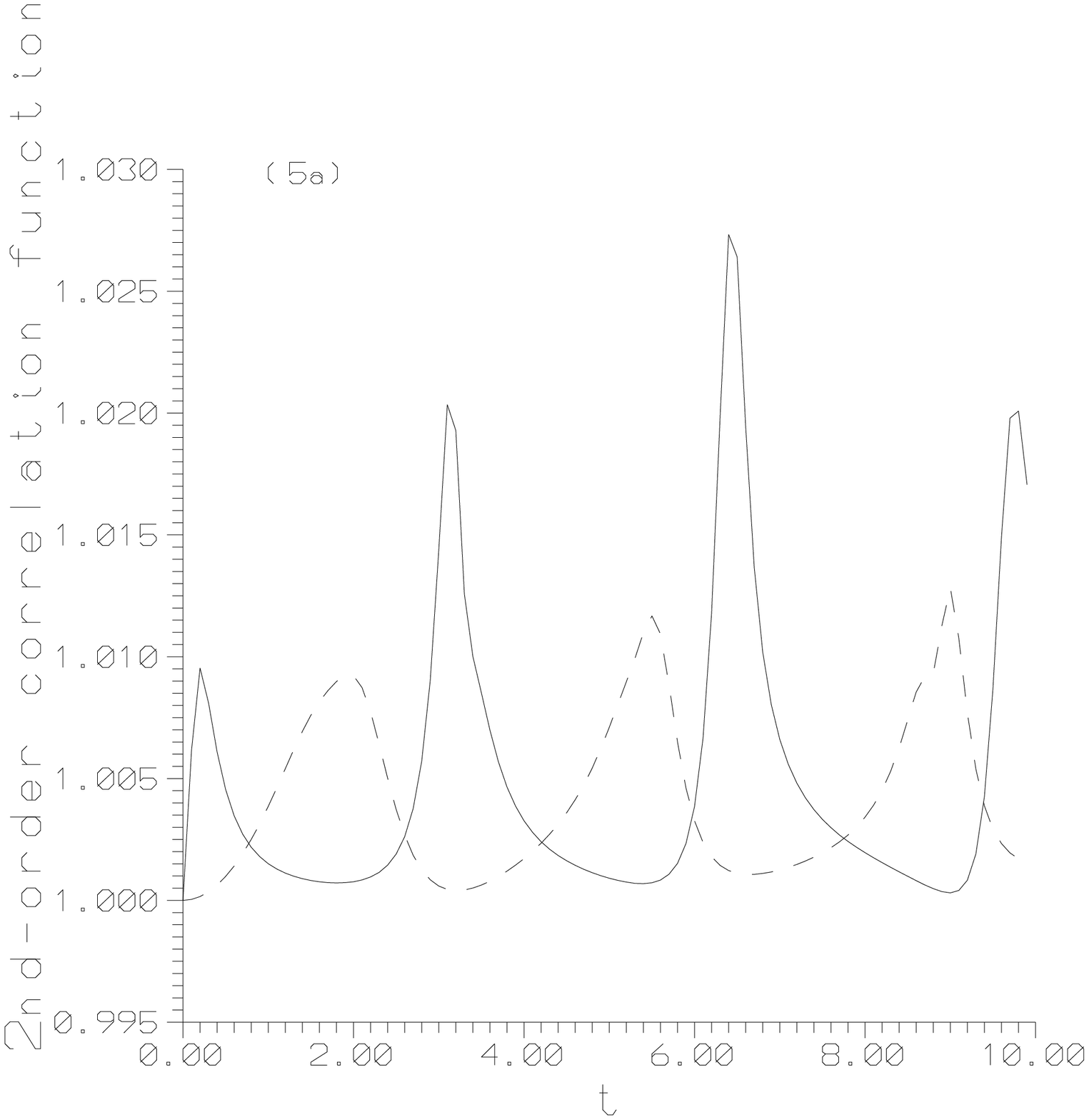}}
 \subfigure[]{\includegraphics[width=8cm]{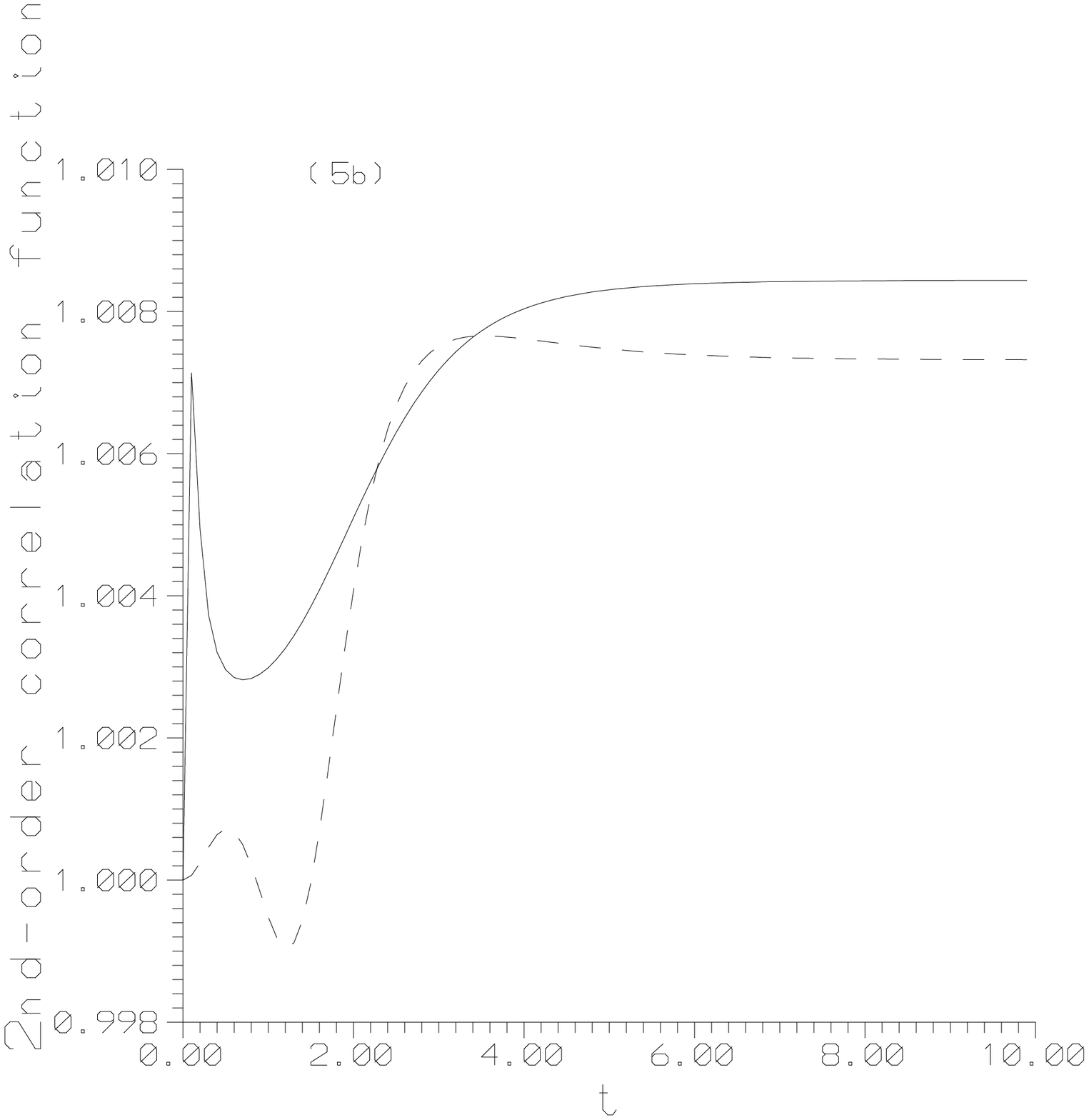}}
    \caption{
Normalized normal second-order correlation function $%
g^{(2)}_{1}(t)$ for mode 1 when both the modes are initially in
the coherent states with $\alpha=5, \beta=20$ (solid curve) and
$\alpha=20, \beta=5$ (dashed curve): a) for both curves
$\lambda_{1}=0.17$,
$\lambda_{2}=0.2$, $%
\lambda_{3}=1$ and $\lambda_{4}=0.2$; b) for both curves
$\lambda_{1}=0.17$, $%
\lambda=0.2$, $\lambda_{3}=1$ and $\lambda_{4}=2$.  }
  \label{fig6}
\end{figure}
A similar behaviour is expectable
for the normalized normal second-order correlation
function for mode 1 if initially both the modes are in coherent
states (Figs. 5 for shown values of the parameters).
In other words, for $\lambda_{3}>\lambda_{4}$,
we observe that $g^{(2)}_{1}(t)$ has oscillatory behaviour between
Poissonian and super-Poissonian statistics, i.e. coherent light can be
approximately recovered at certain values of time.
This behaviour is independent of
the initial amplitudes of the input light (compare solid and
dashed curves). On the other hand, for
$\lambda_{3}<\lambda_{4}$, the oscillatory behaviour disappears and the
fields begin to be localized in the waveguides into which they were initially
launched. The interesting point, which could be realized here,
is that there is
a possibility to generate
 sub-Poissonian light from the initial Poissonian light input into
the coupler provided
that $\alpha >\beta$ (Fig. 5b).

\begin{figure}[h]%
  \centering
  \subfigure[]{\includegraphics[width=8cm]{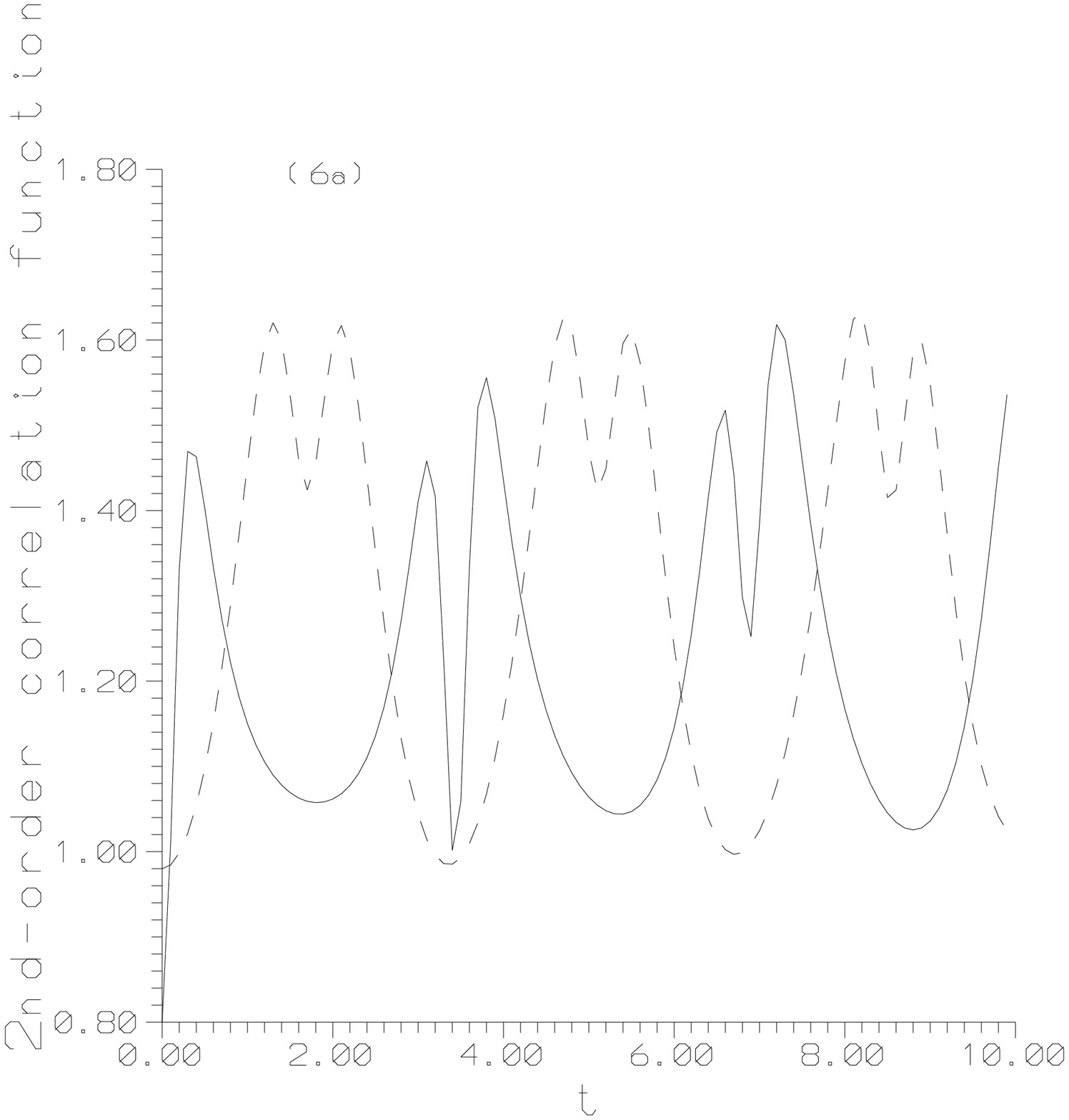}}
 \subfigure[]{\includegraphics[width=8cm]{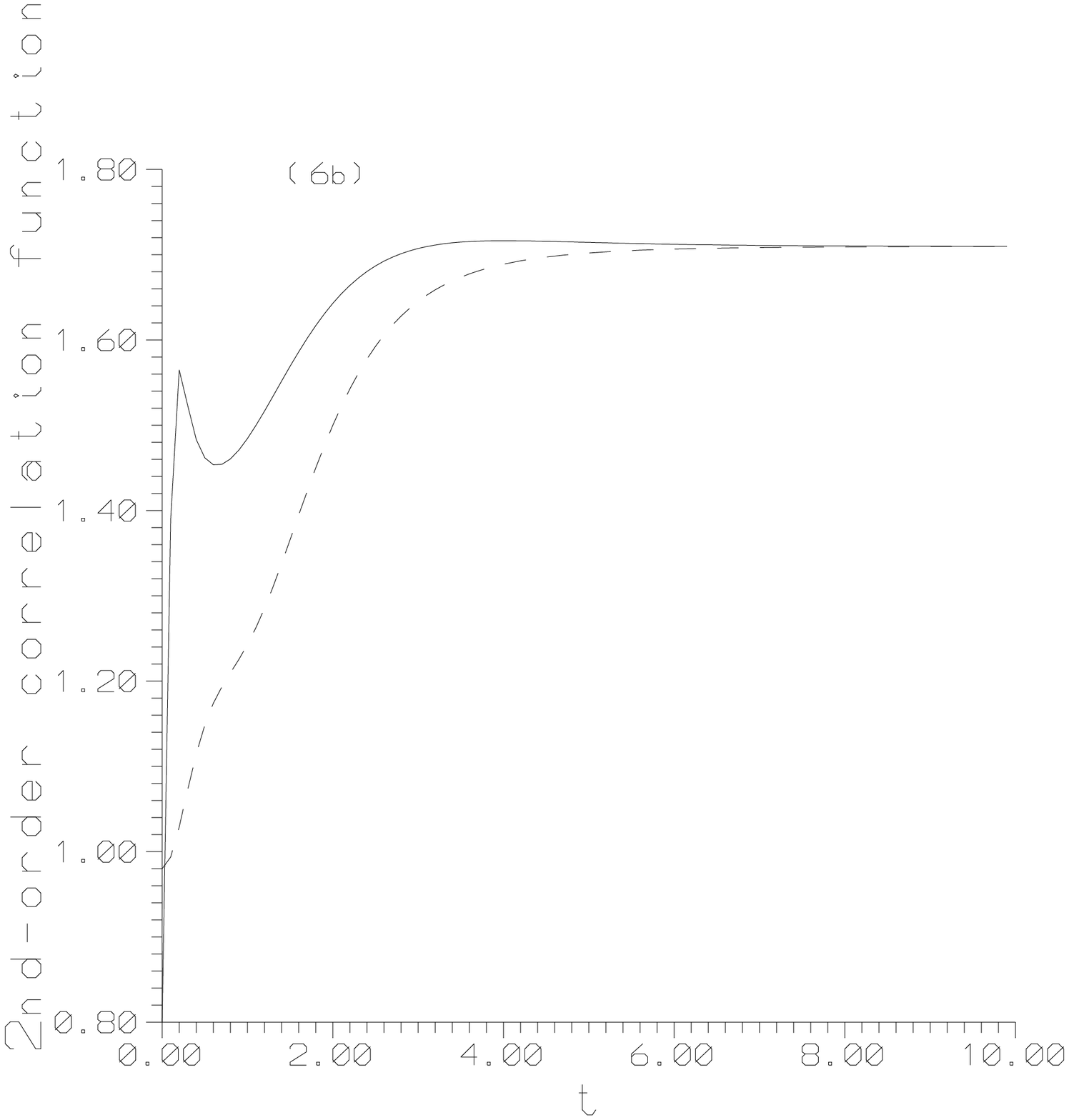}}
    \caption{
Normalized normal second-order correlation function $%
g^{(2)}_{1}(t)$ for mode 1 when both the modes are initially in
the number states with $n=5$, $m=50 $ (solid curve) and $n=50$,
$m=5 $ (dashed curve): a) $\lambda_{k}$ have the same values as in
Fig. 5a; b) $\lambda_{k}$ have the same values as in Fig. 5b.
  }
  \label{fig7}
\end{figure}
The situation will be quite different if we inject initially number
states in the coupler, as is illustrated in Figs. 6,
where we see that the initial
sub-Poissonian statistics for Fock state are not
recovered in the progress of
time $t$ and super-Poissonian statistics dominate.  However, $g_{1}^{(2)}(t)$
exhibits oscillatory behaviour under the condition provided that
the linear coupling is stronger than the nonlinear coupling (Fig. 6a).

We can conclude that this structure can be used to generate nonclassical
light from classical light, e.g. coherent light, by controlling the
device design and the initial input field. Of course, this is based on the
fact that when electromagnetic fields are guided inside the structure,
exchange of energy between the two waveguides is possible because of the
evanescent field between the waveguides \cite{[36]}.

\section{ Quasiprobability functions}

Here we shall continue in our investigation for the statistical
properties of the system under discussion in the basis of
 quasiprobability distribution functions for compound modes
when both the modes are initially
in number,  coherent and  thermal states.

 There are three types of these functions:
Wigner W-, Husimi Q- and Glauber P-functions.
These functions give a complete
description for the statistical properties
of a microscopic
system and provide insight into the nonclassical features of the radiation
fields. For example, the density operator for the quantum mechanical system
can be expressed in terms of them
 and the various moments
of the system operators may be obtained by
appropriate integration in phase space using these functions \cite{[16]}.
Furthermore, these quasidistributions can been determined in
homodyne tomography \cite{leonh}.

On the other hand, as we have mentioned before, propagation of waves inside
the nonlinear directional coupler causes energy exchange between the
waveguides owing to the evanescent waves and hence  if the
measurement of an observable in the first waveguide is performed, this
projects the state of the other waveguide into a new state;
so it would be  convenient  to consider in our investigation
not only the joint quasiprobability  functions but also these
functions for single modes.

The starting point for our analysis is the s-parametrized characteristic
function which is complex in its nature and may be used also to generate the
different moments of the quantum system by means of differentiation.
The two-mode $s$-parametrized characteristic function is given by

$ {\displaystyle  C^{(2)}(\zeta_{1},\zeta_{2}, s, t )= {\rm Tr}\left\{
\hat{\rho} (0)\exp \left[\sum_{i=1}^{2}\left(\frac{s}{2} |\zeta_{i} |^{2} +\zeta_{i}
\hat{a}_{i}^{\dagger}(t) -\zeta_{i}^{*}\hat{a}_{i}(t)\right)\right]
\right\}, }\hfill (26) $

\noindent where $s$ takes on values $1, 0$ and $-1$ corresponding to normally,
symmetrically and antinormally ordered characteristic functions,
respectively, $\hat{\rho}(0)$ is the initial density operator for the model
and {\rm Tr} denotes trace of the operator.

The $s$-parametrized quasiprobability distribution functions are defined
as the Fourier transform of the $s$-parametrized characteristic function
by

$ {\displaystyle  W^{(2)}(\alpha_{1},\alpha_{2},s,t)=\frac{1}{\pi ^{4}}\int
\int d^{2}\zeta_{1}d^{2}\zeta_{2} C^{(2)}(\zeta_{1},\zeta_{2}, s,t) \exp\left[
\sum_{i=1}^{2}(\alpha_{i} \zeta_{i}^{*}-\alpha_{i}^{*}\zeta_{i})\right], }
\hfill (27) $

\noindent where $C^{(2)}(\zeta_{1},\zeta_{2}, s, t )$ is given by (26). When
$s=1, 0,-1$, equation (27) gives formally  $P$- ,
$W$- and  $Q$-functions, respectively.

The corresponding single-mode $s$-parametrized characteristic and
quasiprobability functions are

$ {\displaystyle C^{(1)}(\zeta_{j} ,s,t )= {\rm Tr}\left\{\hat{\rho}(0)
\exp\left[\frac{s }{2} |\zeta_{j} |^{2} +\zeta_{j} \hat{a}_{j}^{\dagger}(t)
-\zeta_{j}^{*}\hat{a}_{j}(t) \right] \right\}, }\hfill (28) $

$ {\displaystyle W^{(1)}(\alpha_{j}, s,t)=\frac{1}{\pi ^{2}}\int
d^{2}\zeta_{j} C^{(1)}(\zeta_{j}, s,t) \exp (\alpha_{j}
\zeta_{j}^{*}-\zeta_{j}\alpha_{j}^{*}), \quad j=1,2. }\hfill (29) $

\noindent The superscripts (1) and (2) in the above equations stand for single-mode
case and two-mode case, respectively.

The various moments of the bosonic operators for
the system, using the characteristic functions and quasiprobability
functions, in the normal
form (N), antinormal form (A) and symmetrical form (S), corresponding to
$s=1,-1,0$, respectively,
 can be obtained by

${\displaystyle
\langle \prod _{j=1}^{2}\hat{a}_{j}^{\dagger m_{j}}(t)\hat{a}_{j}^{n_{j}}(t)
\rangle_{N,A,S}= \prod _{j=1}^{2}\frac{\partial^{m_{j}+n_{j}}}
                           {\partial \zeta _{j}^{m_{j}}
                            \partial (-\zeta _{j}^{*})^{n_{j}}}C^{(2)}
                             (\underline{\zeta },s,t)_{s=1,-1,0}
                             |_{\underline{\zeta }=\underline{\zeta}^{*}=0}}$

${\displaystyle
=\int W^{(2)}(\underline{\alpha },s,t)_{(s=1,-1,0)}\prod _{j=1}^{2}\alpha_{j}^
{*m_{j}}
                         \alpha _{j}^{n_{j}}d^{2}\alpha _{j},}
 \hfill (30)$

\noindent where $n_{j}, m_{j}$ are positive integers,
$\underline{\zeta}=(\zeta_{1},\zeta_{2})$,
$\underline{\alpha}=(\alpha_1,\alpha_2)$,
and the integral is taken over
$\alpha_{1},\alpha_{2}$ in phase
space. For example, when $n_{1}=m_{1}=1$ and $n_{2}=m_{2}=0$, then
 $\langle \hat{a}^{\dagger}_{1}(t)\hat{a}_{1}(t)\rangle_{N}=
\langle \hat{a}^{\dagger}_{1}(t)\hat{a}_{1}(t)\rangle,
\langle \hat{a}^{\dagger}_{1}(t)\hat{a}_{1}(t)\rangle_{A}
=\langle \hat{a}_{1}(t)\hat{a}^{\dagger}_{1}(t)\rangle$,
and
$\langle \hat{a}^{\dagger}_{1}(t)\hat{a}_{1}(t)\rangle_{S}\\
=\frac{1}{2}\langle \hat{a}^{\dagger}_{1}(t)\hat{a}_{1}(t)
+ \hat{a}_{1}(t)\hat{a}^{\dagger}_{1}(t)\rangle$. The formula
(30) is valid for the single and compound modes owing to the normalization
of quasiprobability functions and taking into account that the
single mode
characteristic function can be obtain from that for two modes by simply
setting one of the parameters ($\zeta_{1}$ or $\zeta_{2}$) equals zero.
\newline
{\bf (i) Input Fock states}
\newline
It is known that the nonlinear directional coupler is an important
optical device to generate nonclassical light in the context of
control of light in the nonlinear medium. So the initial input
light has a direct relation to the output light. In fact,
investigation of output light from the coupler, when the number
states are initially injected \cite{[35],buz1,chef1,abdf}, took
little attention compared with the injected coherent states. This
seems to be related to the complexity of calculations. However,
some interesting results have been extracted by considering such
situation \cite{buz1,abdf}.  For example, we can mention, in the
linear directional
 coupler, displaced number states can be generated if a number state enters
 waveguide 1  and a strong coherent field enters waveguide 2 \cite{buz1};
also a coherent
 state has been obtained in the
 nondegenerate optical parametric
symmetric coupler
 when one of the modes enters the
coupler
in the Fock state $|1\rangle$ and the other modes are in vacuum states
 \cite{abdf}.
Here we shall turn our attention  to deduce the quasiprobability functions
for the Hamiltonian (1) when the two modes are initially uncorrelated
and enter the coupler in number states.  Of course, this will give general
formulas having wide applicability for special cases
 \cite{[6],moll1,mish,martin} by appropriate
choice of the parameters. It is important to mention that some of
these special cases have not been considered before
\cite{[6],moll1,mish,martin}.

The density operator for two-mode number states is

$ {\displaystyle \hat{\rho}_{n}(0)={\rm |n\rangle_{1}|m\rangle_{2}}
{\rm_{2}\langle m| _{1}\langle n|}. }\hfill (31) $

\noindent Inserting (31) into (26), the two-mode s-parametrized
characteristic function takes the form

$ {\displaystyle C^{(2)}_{n,m}(\zeta_{1},\zeta_{2},s,t) = \exp \left[ \frac{s
}{2}\left(|\zeta_{1}|^{2}+ |\zeta_{2}|^{2}\right)- \frac{1}{2}\left(
|\eta_{1}(t)|^{2}+|
\eta_{2}(t)|^{2}\right)\right] }\hfill $

$ {\displaystyle \qquad \times L_{n}(|\eta_{1}(t)|^{2})
L_{m}(|\eta_{2}(t)|^{2}), } \hfill (32) $

\noindent where

$ {\displaystyle
\eta_{1}(t)=\zeta_{1}K_{1}^{*}(t)-\zeta^{*}_{1}L_{1}(t)
+\zeta_{2}M_{2}^{*}(t)-\zeta^{*}_{2} N_{2}(t), }\hfill (33a) $

$ {\displaystyle
\eta_{2}(t)=\zeta_{1}M_{1}^{*}(t)-\zeta^{*}_{1}N_{1}(t)
+\zeta_{2}K_{2}^{*}(t)-\zeta^{*}_{2} L_{2}(t), }\hfill (33b) $

\noindent and $L_{n}$ represents the Laguerre polynomial.

Equations (32) and (27) yield Wigner function for two-mode number states;
after some manipulations, we have the following expression

$ {\displaystyle
W^{(2)}_{n,m}(\alpha_{1},\alpha_{2},s=0,t) = \frac{4}{{\pi}^{2}}
(-1)^{(n+m)}  L_{n}(4|\Lambda_{1}(t)|^{2}) L_{m}(4|\Lambda_{2}(t)|^{2}) }
\hfill $

$ {\displaystyle \qquad \times \exp \left[ -2\left(|\Lambda_{1}(t)|^{2}+
|\Lambda_{2}(t)|^{2}\right)\right], } \hfill (34) $

\noindent where

$ {\displaystyle \Lambda_{1}(t)=\alpha_{1}K_{1}^{*}(t)-
\alpha^{*}_{1}L_{1}(t) +\alpha_{2}M_{2}^{*}(t)-\alpha^{*}_{2} N_{2}(t), }
\hfill (35a) $

$ {\displaystyle \Lambda_{2}(t)=\alpha_{1}M_{1}^{*}(t)-
\alpha^{*}_{1}N_{1}(t) +\alpha_{2}K_{2}^{*}(t)-\alpha^{*}_{2} L_{2}(t). }
\hfill (35b) $

\noindent Equation (34) cannot be factorized owing to the intermodal
correlation between the propagating modes inside the coupler and this is
clear since (34) includes terms like $\alpha_{1}\alpha_{2},\alpha^{*}_{1}
\alpha_{2}$, etc.

The single-mode s-parametrized characteristic function
for the first mode can be obtained, by
similar way as for the two-mode case, from (28) as

$ {\displaystyle
C^{(1)}_{n,m}(\zeta_{1},s,t)=\exp \left[ \frac{s}{2}|\zeta_{1}|^{2}
-\frac{1}{2}\left(|\nu_{1}(t)|^{2}+|\nu_{2}(t)|^{2}\right)\right]
L_{n}(|\nu_{1}(t)|^{2})
L_{m}(|\nu_{2}(t)|^{2}), }\hfill (36) $

\noindent where

$ {\displaystyle
\nu_{1}(t)=\zeta_{1}K_{1}^{*}(t)-\zeta^{*}_{1}L_{1}(t), }
\hfill (37a) $

$ {\displaystyle
\nu_{2}(t)=\zeta_{1}M_{1}^{*}(t)-\zeta^{*}_{1}N_{1}(t). }
\hfill (37b) $

Inserting (36) into (29), carrying out the integration and taking $s=0$
and $s=-1$, the W-function and Q-function for the single-mode case can be
obtained:

$ {\displaystyle
W^{(1)}_{n,m}(\alpha_{1},s,t) = \frac{2(n!m!)
\exp \left[\frac{(\alpha_{1}e^{-\frac{i\epsilon(t)}{
2}} -\alpha^{*}_{1}e^{\frac{i\epsilon(t)}{2}} )^{2}} {2(\tau(t)
-s-2|\psi(t)|)} -\frac{(\alpha_{1}e^{-\frac{i\epsilon(t)}{2}}
+\alpha^{*}_{1}e^{\frac{i\epsilon(t)}{2}} )^{2}} {2(\tau(t)-s+2|\psi(t)|)}
\right] }{\pi\sqrt{(\tau(t)-s)^{2}-4|\psi(t)|^{2}}} }\hfill $

$ {\displaystyle \qquad \times
\sum_{l_{1}=0}^{n}\sum_{n_{1}=0}^{l_{1}}\sum_{l_{2}=0}^{n}
\sum_{m_{1}=0}^{l_{2}}\sum_{k_{1}=0}^{m+l_{2}-n}
\sum_{k_{2}=0}^{m+l_{1}-n}
(-2)^{2n-2r+m_{1}+n_{1}-l_{1}-l_{2}} }\hfill $

$\hfill$

$ {\displaystyle \qquad \times\frac{
r!(m-n+m_{1}+l_{2}-k_{1}-r)![n_{1}!m_{1}!(m-n+l_{2}-k_{1})! (m-n+l_{1}-
k_{2})!]^{-\frac{1}{2}}}{k_{1}!k_{2}!(l_{1}-n_{1})!(l_{2}-m_{1})!
(n-l_{1})!(n-l_{2})!} }\hfill $

$\hfill$

$ {\displaystyle \qquad \times [\eta_{+}^{2}(t)-\eta_{-}^{2}(t)
]^{\frac{
l_{1}-n_{1}}{2}} [\eta_{+}^{*2}(t)- \eta_{-}^{*2}(t)
]^{\frac{l_{2}-m_{1}}{2}
} [\zeta_{+}^{2}(t)-\zeta_{-}^{2}(t) ]^{\frac{k_{1}}{2}}
[\zeta_{+}^{*2}(t)-\zeta_{-}^{*2}(t) ]^{\frac{k_{2}}{2}} }\hfill $

$\hfill$

$ {\displaystyle \qquad \times
[\eta_{-}(t)\zeta_{-}^{*}(t)+\eta_{+}(t)\zeta_{+}^{*}(t) ]^{n-l_{1}}
[\eta_{-}^{*}(t)\zeta_{-}(t)+\eta_{+}^{*}(t)\zeta_{+}(t) ]^{n-l_{2}}
}\hfill $

$\hfill$

$ {\displaystyle \qquad \times (\frac{\eta_{+}^{*}(t)\zeta_{+}^{*}(t)-
\eta_{-}^{*}(t)\zeta_{-}^{*}(t) }{z(t)} )^{m_{1}-r} (\frac{
\eta_{+}(t)\zeta_{+}(t)-\eta_{-}(t)\zeta_{-}(t)}{z(t)} )^{n_{1}-r}
}\hfill $

$\hfill$

$ {\displaystyle \quad \times
[1-2(|\zeta_{+}(t)|^{2}+|\zeta_{-}(t)|^{2}
)]^{\frac{2m-2n+l_{1} +l_{2}-k_{1}-k_{2}}{2}} }\hfill $

$\hfill$

${\displaystyle \qquad \times [1-2(|\eta_{+}(t)|^{2}+|\eta_{-}(t)|^{2}
)]^{\frac{n_{1}+m_{1}}{2}} } $

$\hfill$

${\displaystyle
 \displaystyle \times [1- \frac{4}{z^{2}(t)}|\zeta_{-}(t)\eta_{-}(t)-
\zeta_{+}(t) \eta_{+}(t)|^{2}]^{r}}$

$\hfill$

$ {\displaystyle \qquad \times H_{l_{1}-n_{1}}\left(\frac{X(t)}{\sqrt{
\eta^{2}_{+}(t)- \eta^{2}_{-}(t)}}\right)
H_{l_{2}-m_{1}}\left(\frac{X^{*}(t)
}{\sqrt{ \eta^{*2}_{+}(t)- \eta^{*2}_{-}(t)}} \right) }\hfill $

$\hfill$

$ {\displaystyle \qquad \times H_{k_{1}}\left(\frac{Y(t)}{\sqrt{
\zeta^{2}_{+}(t)- \zeta^{2}_{-}(t)}}\right)
H_{k_{2}}\left(\frac{Y^{*}(t)}{
\sqrt{\zeta^{*2}_{+}(t) -\zeta^{*2}_{-}(t)}}\right) }\hfill $

$\hfill $

$ {\displaystyle \qquad \times
P_{r}^{(|m_{1}-n_{1}|,|n_{1}+n+k_{1}-m-l_{2}|)}\left[ \frac{
z^{2}(t)-4|\zeta_{-}(t)\eta_{-}(t)-\zeta_{+}(t)\eta_{+}(t)|^{2}}{
z^{2}(t)+4|\zeta_{-}(t)\eta_{-}(t)-\zeta_{+}(t)\eta_{+}(t)|^{2}}\right],}
\hfill (38) $

\noindent where

$ {\displaystyle |\tau(t)|^{2}=|K_{1}(t)|^{2}+|L_{1}(t)|^{2}
+|M_{1}(t)|^{2}+|N_{1}(t)|^{2}, }\hfill (39a) $

$ {\displaystyle \psi(t)=K_{1}(t)L_{1}(t)+N_{1}(t)M_{1}(t)
=|\psi(t)|e^{i\epsilon(t)}, }\hfill (39b) $

$ {\displaystyle
\eta_{\pm}(t)=\frac{K^{*}_{1}(t)e^{\frac{i\epsilon(t)}{2}}
\pm L_{1}(t)e^{\frac{-i\epsilon(t)}{2}} } {\sqrt{2(\tau(t) -s\pm
2|\psi(t)|)}
}, }\hfill (39c) $

$ {\displaystyle
\zeta_{\pm}(t)=\frac{M^{*}_{1}(t)e^{\frac{i\epsilon(t)}{2}}
\pm N_{1}(t)e^{\frac{-i\epsilon(t)}{2}} } {\sqrt{2(\tau(t) -s\pm
2|\psi(t)|)}%
}, }\hfill (39d) $

$ {\displaystyle
X(t)=\frac{\eta_{+}(t)(\alpha_{1}e^{-\frac{i\epsilon(t)}{2}
} +\alpha^{*}_{1}e^{\frac{i\epsilon(t)}{2}} )} {\sqrt{2(\tau(t)
-s+2|\psi(t)|)}} +\frac{\eta_{-}(t)(\alpha_{1}e^{-\frac{i\epsilon(t)}{2}}
-\alpha^{*}_{1}e^{\frac{i\epsilon(t)}{2}} )} {\sqrt{2(\tau(t)-s-2|\psi(t)|)}}
, }\hfill (39e) $

$ {\displaystyle
Y(t)=\frac{\zeta_{+}(t)(\alpha_{1}e^{-\frac{i\epsilon(t)}{2}
} +\alpha^{*}_{1}e^{\frac{i\epsilon(t)}{2}} )} {\sqrt{2(\tau(t)
-s+2|\psi(t)|)}}
+\frac{\zeta_{-}(t)(\alpha_{1}e^{-\frac{i\epsilon(t)}{2}}
-\alpha^{*}_{1}e^{\frac{i\epsilon(t)}{2}} )}
{\sqrt{2(\tau(t)-s-2|\psi(t)|)}}
, }\hfill (39f) $

$ {\displaystyle
z(t) =  \sqrt{[1-2(|\eta_{+}(t)|^{2}+|\eta_{-}(t)|^{2})]
[1-2(|\zeta_{+}(t)|^{2}+|\zeta_{-}(t)|^{2})]}, }\hfill (39g) $
$r=\frac{1}{2}[n_{1}+m_{1}-|n_{1}-m_{1}|]$, $H_{m}$ is the Hermite
polynomial of order $m$ and $P^{(a,b)}_{r}(x)$ is the Jacobi polynomial
which is defined as

$ {\displaystyle P_{r}^{(c,d)}(x) = \sum_{k=0}^{r}
(-1)^{(r-k)}{r+d \choose
r-k} {r+k+c+d \choose k } (\frac{x+1}{2})^{k}. }\hfill (40) $

Equation (38) is real in spite of its complex form, which can be seen
explicitly in the summations where we can find each term with its complex
conjugate.

We can easily check the limits of equations (36) and (38) as
$t\rightarrow 0$, which give the corresponding well-known quantities for the
Fock state $|n\rangle$ appropriate for the description before the interaction
starts.
In fact, this is clear also from the solutions of the Heisenberg equations of
motion, where at $t=0$ the all factors reduce to zero except $K_{1}(0)$
which equals 1. So we get

$ {\displaystyle C^{(1)}(\zeta ,s)=\exp [\frac{1}{2}(s-|\zeta|^{2})]
L_{n}(|\zeta|^{2}), }\hfill (41) $

$ {\displaystyle W(\alpha)=\frac{2}{\pi}(-1)^{n}
\exp(-2|\alpha|^{2})L_{n}(4|\alpha|^{2}), }\hfill (42) $

$ {\displaystyle Q(\alpha)=\frac{1}{\pi}\frac{|\alpha|^{2n}}{n!}
\exp(-|\alpha|^{2}),}\hfill (43) $

\noindent which are the $s$-parametrized characteristic function, W-function and
Q-function for the Fock state $| n\rangle$.

As we have mentioned before, the nonlinear directional coupler can be used
as a source of quantum states \cite{abdf}. This may be illustrated
by displaying one of the
quasiprobability functions \cite{gill}.
The best quasiprobability functions for this task are  $W$- and $Q$-functions
since they are not singular and may contain oscillatory fringes
(particularly $W$-function) that are
indicative of nonclassical behaviours.
\begin{figure}[h]%
  \centering
  \subfigure[]{\includegraphics[width=8cm]{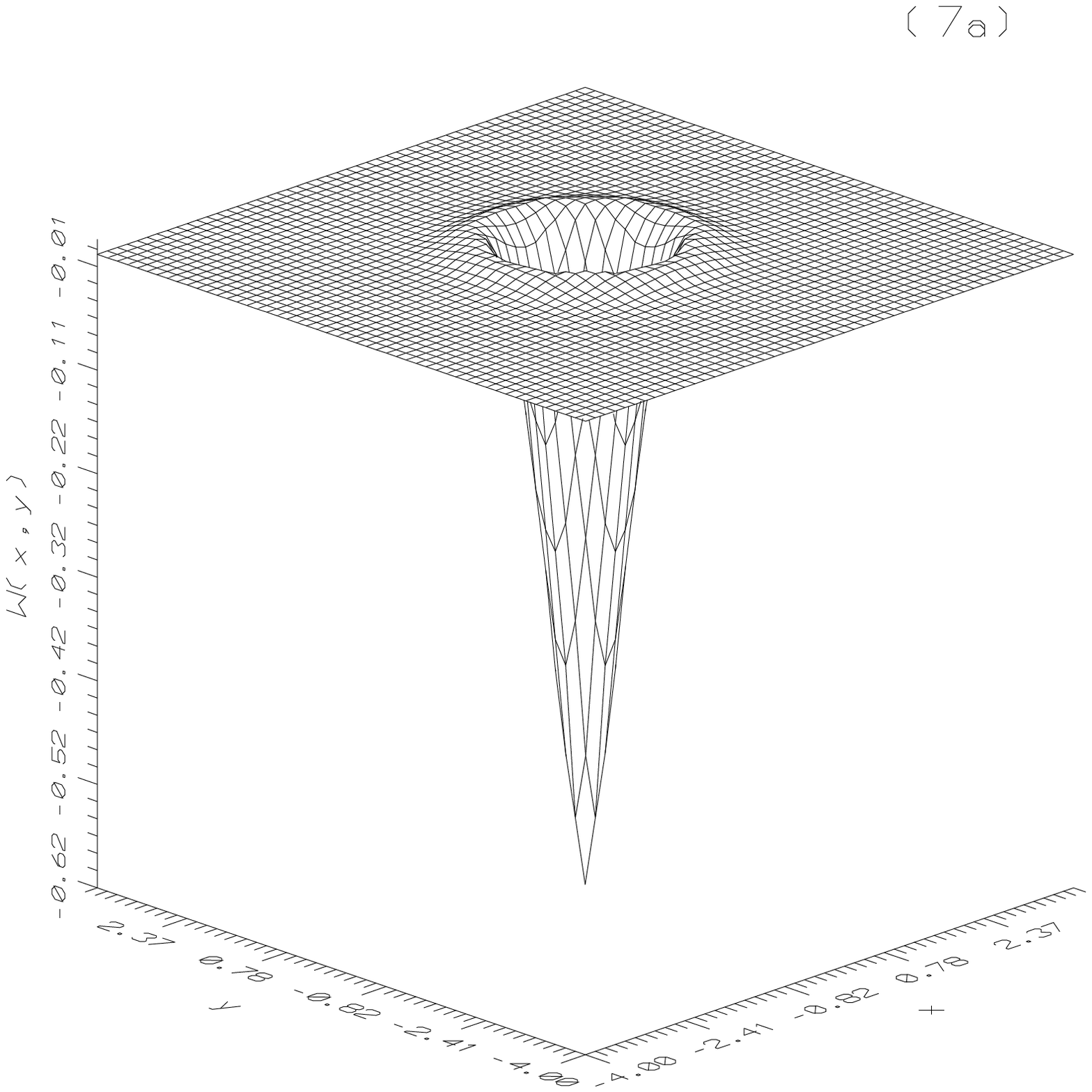}}
 \subfigure[]{\includegraphics[width=8cm]{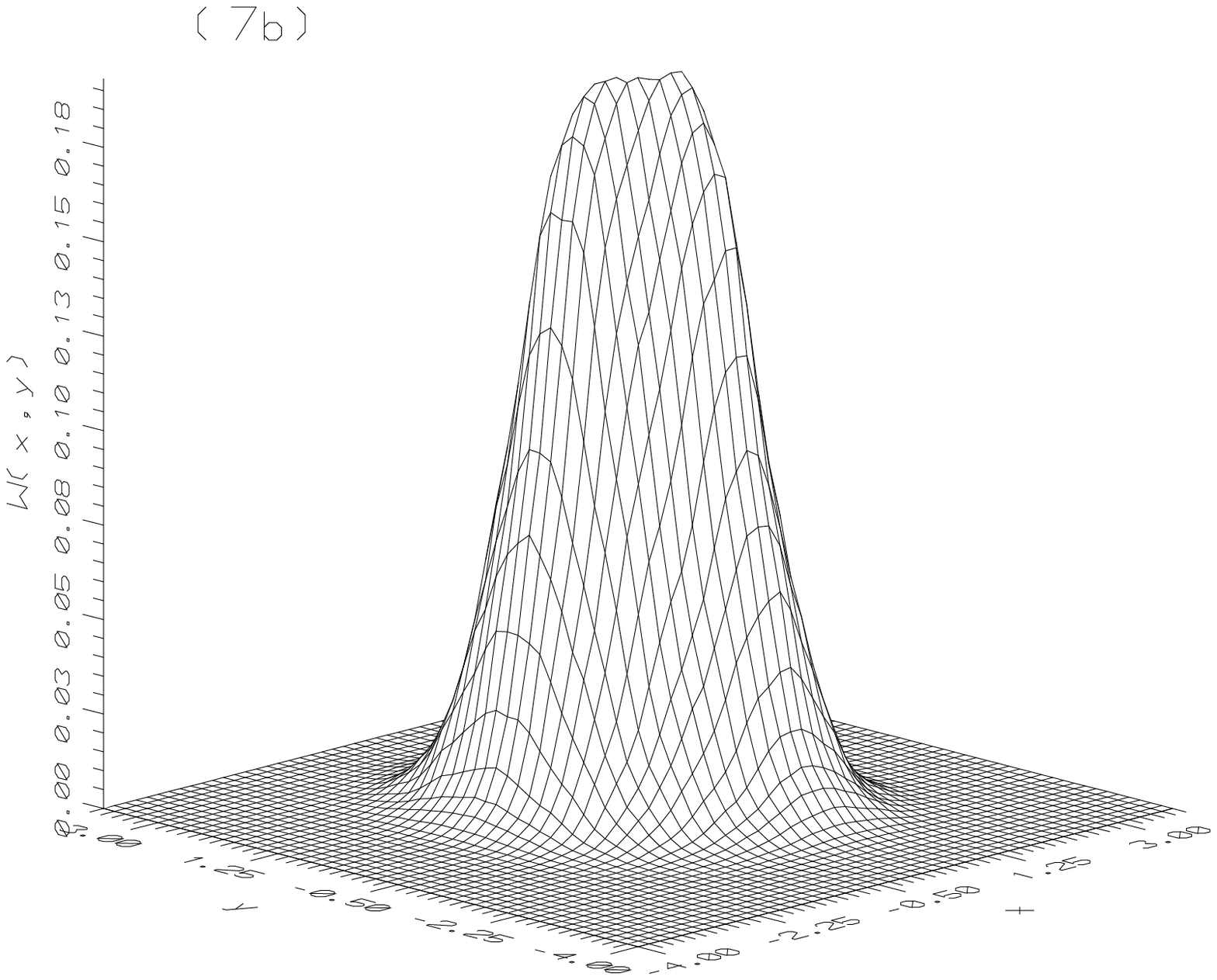}}
 \subfigure[]{\includegraphics[width=8cm]{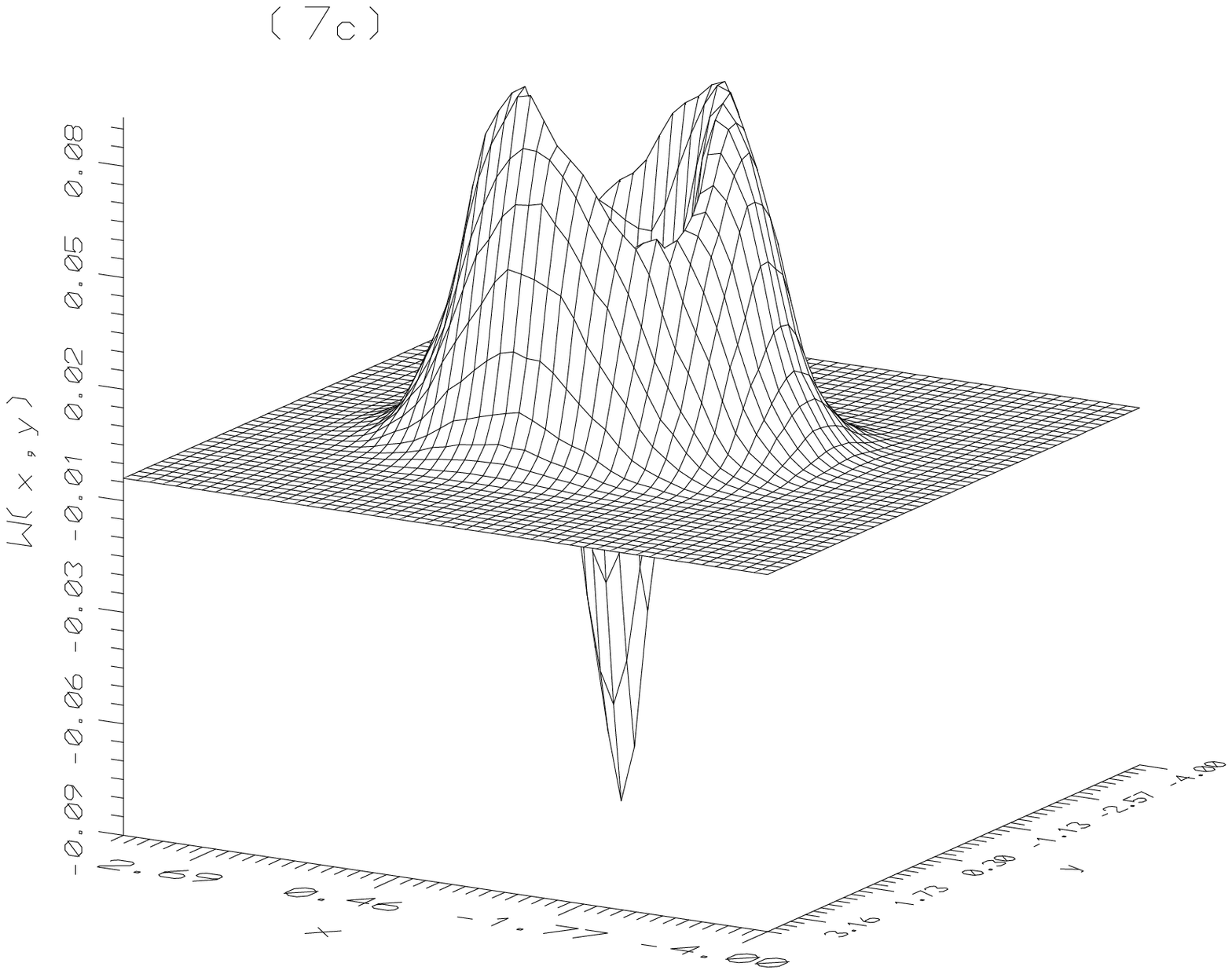}}
    \caption{
W-function for the single mode (mode 1) for different values of
time $t$ when both the modes are initially in the number states;
the first mode is in the state $|1\rangle $ and the second mode is
in the
state $%
|0\rangle$ and $\lambda_{3}=1$,
$\lambda_{1}=\lambda_{2}=\lambda_{4}=0.25$: a) for $t=\pi/100$; b)
for $t=\pi/2$; c) for $t=\pi$.
  }
  \label{fig8}
\end{figure}
So we have plotted W-function and Q-function using (38) in Figs. 7
and 8, respectively, against $x={\rm Re}\alpha_{1}$ and $y={\rm
Im}\alpha_{1}$, when the first state is the Fock state $|1\rangle$
and the second one is the vacuum state $|0\rangle$, i.e. $n=1,
m=0$; $\lambda_{3}=1, \lambda_{1}=\lambda_{2}=\lambda_{4}=0.25$
and for shown values of time. We have considered quasiprobability
functions at $t\in [0,\pi]$. In Fig. 7a we have the W-function for
$t=\frac{\pi}{100}$, i.e. after short time interaction between the
two modes we observe similar behaviour as for the W-function of
the state $|1\rangle$ (see Fig. 2 of \cite{buz2}), which means
that pronounced negative values are exhibited. This behaviour of
the W-function is completely different by increasing the time
($t=\frac{\pi}{2}$); we see disappearance of negative values of
the quasidistribution and a stretched positive peak occurs (Fig.
7b). This form of $W$-function is close to that of squeezed vacuum
states \cite{yun1}, i.e. squeezed vacuum states can be generated,
in principle, in our model. It should be borne in mind that the
specific direction of stretching for the quasiprobability function
of squeezed states may be achieved by choosing a suitable value
for the  phase of squeezing parameter. Of course, in Fig. 7b,
there is a negligible spike at the top of the peak which can be
smoothed out by governing the coupler parameters. After larger
time interaction $t=\pi$, the negative values are reached again
 but they are less pronounced and
 asymmetry  can be observed due to stretching (Fig. 7c).
So we meet a time development of the W-function as a result of the power
transfer
between the two modes inside the coupler.
\begin{figure}[h]%
  \centering
  \subfigure[]{\includegraphics[width=8cm]{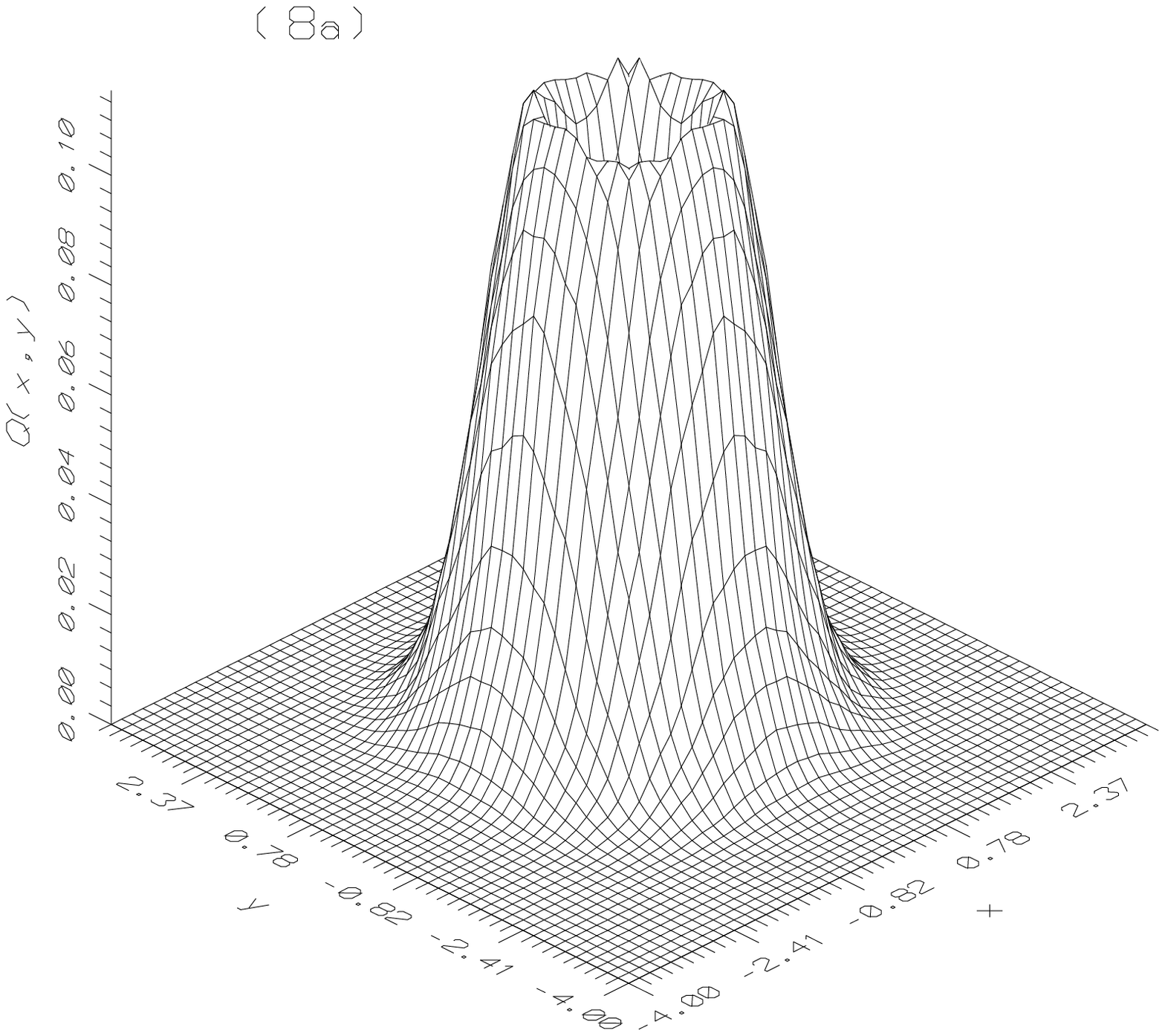}}
 \subfigure[]{\includegraphics[width=8cm]{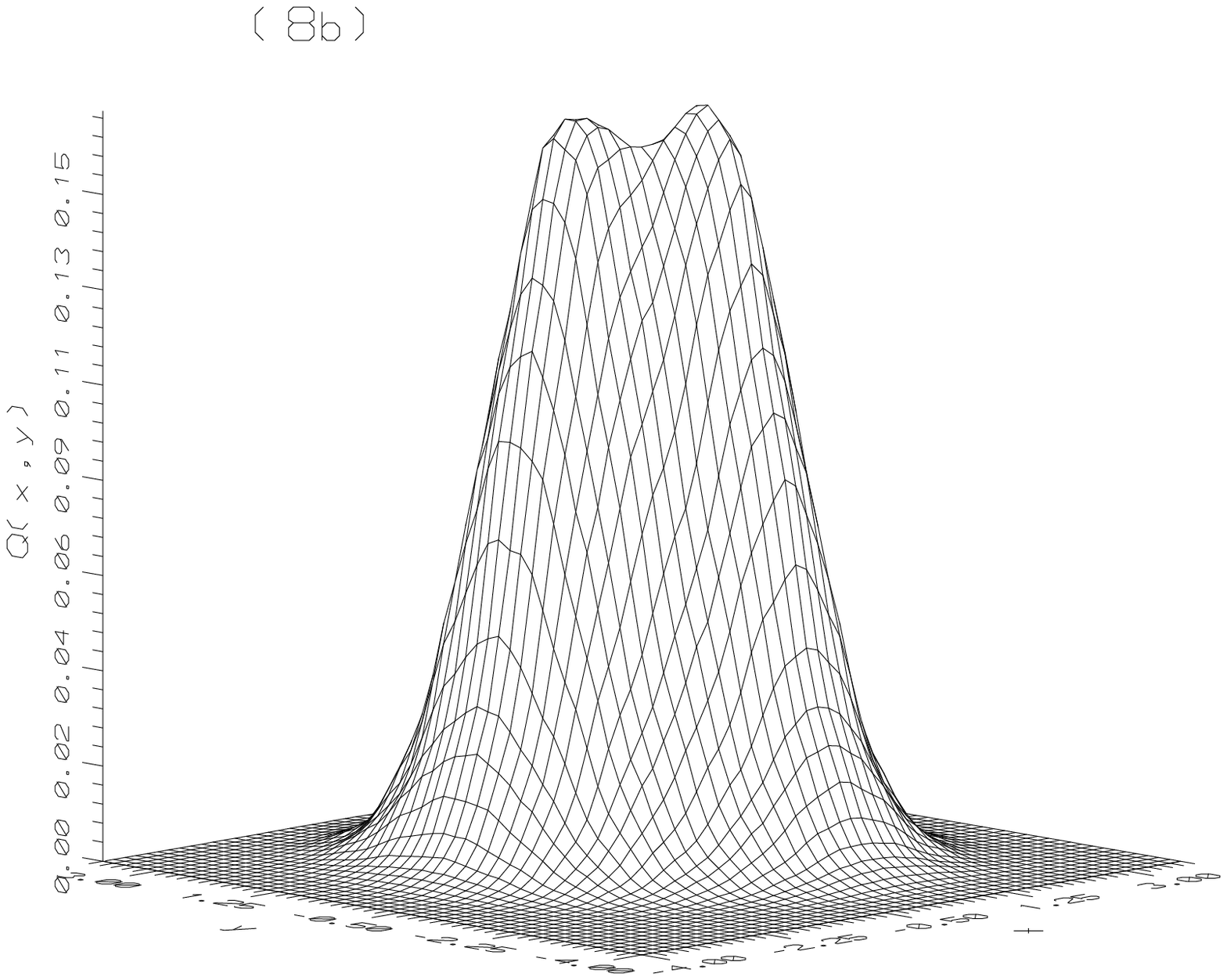}}
 \subfigure[]{\includegraphics[width=8cm]{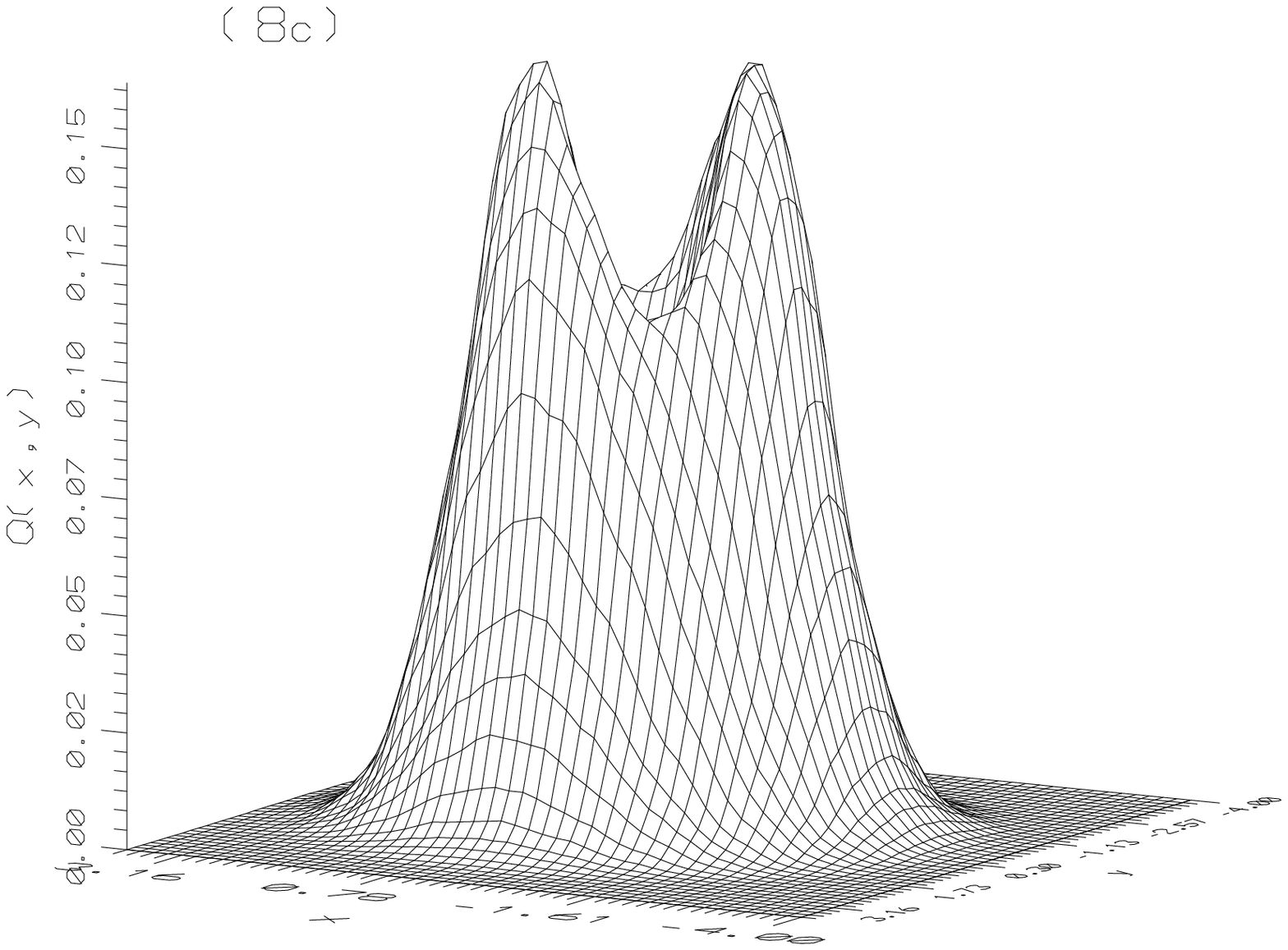}}
    \caption{
Q-function for the single mode (mode 1) for different values of
time $t$ when both the modes are initially in the number states;
the values of parameters are as in Fig. 7.
 }
  \label{fig9}
\end{figure}
The Q-function is the quasiprobability function which is always
positive definite,
however, it can be used as an indicator for the squeezing
in the model by including stretching in the phase space.
In Fig. 8 we can see a kind of relation of the behaviour of W-function
and Q-function and we can observe the top hole peak for short and long
time interaction, which does not appear for intermediate interaction
times. For all cases the stretching is remarkable.
\newline
{\bf (ii) Input coherent light} \newline
In a similar way as we followed in the case {\bf (i)} we can study the same
quantities when both the modes are initially in coherent states. In this
case the density operator is given by

$ {\displaystyle \hat{\rho}_{{\rm coh}}(0)={\rm
|\alpha\rangle_{1}|\alpha
\rangle_{2}}{\rm _{2}\langle \alpha |}{\rm _{1}\langle \alpha |}.}
\hfill (44) $

Then the two-mode $s$-parametrized characteristic function is derived in the
form
$ {\displaystyle C^{(2)}_{{\rm coh}}(\zeta_{1},\zeta_{2},s,t) = \exp
\Bigl\{\sum_{j=1}^{2}\left[- \frac{1}{2}\left(1-s+
2|L_{j}(t)|^{2}+2|N_{j}(t)|^{2}\right)
\right]|\zeta_{j}|^2\Bigr\} }\hfill $

$ {\displaystyle \qquad \times\exp\Bigl\{\sum_{j=1}^{2} \frac{1}{2}\left[
\zeta^{2}_{j}
\left(N^{*}_{j}(t)M^{*}_{j}(t)+L^{*}_{j}(t)K^{*}_{j}(t)\right)+{\rm c.c.}
\right]\Bigr\} }\hfill $

$\hfill$

$ {\displaystyle \qquad \times\exp \Bigl\{\zeta_{1}\zeta_{2}
[K^{*}_{1}(t)N^{*}_{2}(t)+ M^{*}_{1}(t)L^{*}_{2}(t)] + {\rm c.c.}
\Bigr\} }
\hfill $

$\hfill$

$ {\displaystyle \qquad \times\exp \Bigl\{-\zeta^{*}_{1}\zeta_{2}[
L^{*}_{2}(t)N_{1}(t)+ N^{*}_{2}(t)L_{1}(t)]- {\rm c.c.}\Bigr\} }
\hfill $

$\hfill$

$ {\displaystyle \qquad \times
\exp\Bigl\{\sum_{j=1}^{2}\left[\zeta_{j}\bar{\alpha}
_{j}^{*}(t)- \zeta_{j}^{*}\bar{\alpha}(t)\right]\Bigr\},} \hfill(45) $

\noindent where $\bar{\alpha}_{j}(t)$ are the mean values of the operators $
\hat{a}_{j}(t)$ with respect to the coherent states.

Therefore the two-mode $s$-parametrized quasiprobability function is

$ {\displaystyle
W^{(2)}_{{\rm coh}}(\alpha_{1},\alpha_{2},s,t) = \frac{1}{\pi ^{2}}
\left[\left(|L_{1}(t)|^{2}+|N_{1}(t)|^{2}-|B_{1}(t)|^{2}\right)\left(S_{+}(t)
S_{-}(t)-T^{2}(t)\right) \right]^{-\frac{1}{2}}} \hfill $

$\hfill$

$ {\displaystyle \qquad \times \exp \left[\frac{S_{-}(t)X_{+}^{2}(t)
+S_{+}(t)X_{-}^{2}(t)+2X_{- }(t)X_{+}(t)T(t)} {S_{+}(t)S_{-}(t)-T^{2}(t)
}
\right] }\hfill $

$\hfill $

$ {\displaystyle \times\exp\left[ \frac{\frac{|B_{1}(t)|}{2}
[E_{1}^{2}(t)+E_{1}^{*2}(t)]- [|L_{1}(t)|^{2}+|N_{1}(t)|^{2}]|E_{1}(t)|^{2}
} {|L_{1}(t)|^{2}+|N_{1}(t)|^{2}-|B_{1}(t)|^{2}}\right], } \hfill (46) $

\noindent where we have used the following abbreviations

$ {\displaystyle
A_{j}(t)=\frac{1}{2}(1-s+2|L_{j}(t)|^{2}+2|N_{j}(t)|^{2}),
}\hfill $

$ {\displaystyle
B_{j}(t)=N^{*}_{j}(t)M^{*}_{j}(t)+L^{*}_{j}(t)K^{*}_{j}(t)
=|B_{j}(t)|e^{2i\delta_{j}(t)}, }\hfill $

$ {\displaystyle
D(t)=K^{*}_{1}(t)N^{*}_{2}(t)+M^{*}_{1}(t)L^{*}_{2}(t)
=|D(t)| e^{i\chi(t)}, }\hfill $

$ {\displaystyle
\bar C(t)=L^{*}_{2}(t)N_{1}(t)+ N^{*}_{2}(t)L_{1}(t)=|\bar
C(t)|e^{i\gamma (t)}, }\hfill $

$ {\displaystyle E_{j}(t)=(\bar{\alpha_{j}}(t)-\alpha_{j})e^{i
\delta_{j}(t)}, }\hfill $

$ {\displaystyle F_{\pm}(t)=D(t)\sin
[\delta_{1}(t)+\delta_{2}(t)-\chi(t)]\pm \bar C(t)  \sin
[\delta_{1}(t)-\delta_{2}(t)+\gamma (t)], }\hfill $

$ {\displaystyle R_{\pm}(t)=D(t)\cos
[\delta_{1}(t)+\delta_{2}(t)-\chi(t)]\pm \bar C(t) \cos
[\delta_{1}(t)-\delta_{2}(t)+\gamma(t)], }\hfill $

$ {\displaystyle
S_{+}(t)=A_{2}(t)+|B_{2}(t)|-\frac{F_{+}^{2}(t)}{A_{1}(t)
-|B_{1}(t)|} -\frac{R_{+}^{2}(t)}{A_{1}(t)+|B_{1}(t)|}, }\hfill $

$ {\displaystyle S_{-}(t)=A_{2}(t)-|B_{2}(t)|-\frac{R_{-}^{2}(t)} {
A_{1}(t)-|B_{1}(t)|}-\frac{F_{-}^{2}(t)}{A_{1}(t)+|B_{1}(t)|}, }\hfill $

$ {\displaystyle T(t)=\frac{R_{-}(t)F_{+}(t)}{A_{1}(t)-|B_{1}(t)|}
-\frac{
R_{+}(t)F_{-}(t)}{A_{1}(t)+|B_{1}(t)|}, }\hfill $

$ {\displaystyle X_{+}(t)=i[E_{2}(t)+E_{2}^{*}(t)] +\frac{F_{+}(t)
[E_{1}^{*}(t)-E_{1}(t)]}{A_{1}(t)-|B_{1}(t)|}-i\frac{R_{+}(t)
[E_{1}^{*}(t)+E_{1}(t)]}{A_{1}(t)+|B_{1}(t)|}, }\hfill $

$ {\displaystyle X_{-}(t)=[E_{2}^{*}(t)-E_{2}(t)]+\frac{R_{-}(t)
[E_{1}^{*}(t)-E_{1}(t)]}{A_{1}(t)-|B_{1}(t)|}+i\frac{F_{-}(t)
[E_{1}^{*}(t)+E_{1}(t)]}{A_{1}(t)+|B_{1}(t)|}, }\hfill (47) $

\noindent with the following condition $|A_{j}(t)|>|B_{j}(t)|$ for the Glauber
P-function and no additional constrains.

From equation (46) we can see that
$W^{(2)}(\alpha_{1},\alpha_{2},t,s)$ includes the nonclassical
correlation nature due to the presence of the terms
$\alpha_{1}\alpha_{2},\alpha^{*}_{1}\alpha_{2}$, etc. These
mode correlations have been used in a number of studies on nonclassical
aspects of light including questions like violations of
Bell inequalities \cite{aga}.
The amount of correlation between the waveguides inside the coupler is
governed by the coupler parameters, i.e. $\alpha_{j},\lambda_{j},t$.
Further, the $P$-function does not exist for
 $|A_{j}(t)|<|B_{j}(t)|$,
  and this should be reflected as a nonclassical effect in the
behaviour of the compound modes inside the coupler.
The
physical reason for this is that the modes may no longer fluctuate
independently in even small amount allowed in a pure state.

For the single-mode case the $s$-parametrized characteristic function and
the $s$-parametrized quasiprobability function are given, respectively, as

$ {\displaystyle
C^{(1)}_{{\rm coh}}(\zeta,s,t) = \exp \left[- \frac{1}{2}\left(1-s+
2|L_{1}(t)|^{2}+2|N_{1}(t)|^{2}\right)|\zeta|^{2} +\zeta\bar{\alpha}
_{1}^{*}(t)-\zeta^{*}\bar{\alpha}_{1}(t) \right] }\hfill $

$\hfill$

$ {\displaystyle \qquad \times \exp\left\{\frac{1}{2} \zeta^{2}
[N^{*}_{1}(t)M^{*}_{1}(t)+L^{*}_{1}(t)K^{*}_{1}(t)]\right\} }\hfill $

$\hfill$

$ {\displaystyle \qquad \times \exp\left\{ \frac{1}{2}\zeta^{*2}
[N_{1}(t)M_{1}(t)+L_{1}(t)K_{1}(t)] \right\}, }\hfill (48) $

$ {\displaystyle
W^{(1)}_{{\rm coh}}(\alpha,s,t) = \frac{1}{\pi\sqrt{ [\frac{1-s}{2}
+|L_{1}(t)|^{2}+|N_{1}(t)|^{2}]^{2}- |B_{1}(t)|^{2} }}} \hfill $

$\hfill$

$ {\displaystyle
 \qquad \times \exp \left\{-\frac{[\frac{1-s}{2}
+|L_{1}(t)|^{2}+|N_{1}(t)|^{2}] |\bar{\alpha_{1}}-\alpha_{1}|^{2}}
{[\frac{1-s}{2} +|L_{1}(t)|^{2}+|N_{1}(t)|^{2}]^{2}-|B_{1}(t)|^{2}}
\right\}
}\hfill $

$\hfill$

$ {\displaystyle \qquad \times \exp \left\{- \frac{\frac{1}{2}
|B_{1}(t)|[E_{1}^{2}(t)+E_{1}^{*2}(t)]} {[\frac{1-s}{2}
+|L_{1}(t)|^{2}+|N_{1}(t)|^{2}]^{2}-|B_{1}(t)|^{2}} \right\}, }\hfill
(49) $

\noindent and $|L_{1}(t)|^{2}+|N_{1}(t)|^{2}>|B_{1}(t)|$ must hold for
the Glauber P-function.
It is known that the correspondence
 between quantum and classical theories can be established via
the Glauber-Sudarshan $P$-representation.
But the $P$-representation does not possess all the
 properties of a classical distribution function for quantum
 fields.
More precisely,
 light fields for which the $P$-representation
is not a well-behaved distribution (in most processes in
interaction at least for some values of interaction time, including the
process under consideration) exhibit nonclassical features.
From (49), the $P$-function is not well defined as an ordinary function for
$|L_{1}(t)|^{2}+|N_{1}(t)|^{2}<|B_{1}(t)|$ and hence the
nonclassical effects,
e.g. squeezing of vacuum fluctuations and sub-Poissonian statistics
can occur, as we have seen
before.
\begin{figure}[h]%
  \centering
  \subfigure[]{\includegraphics[width=8cm]{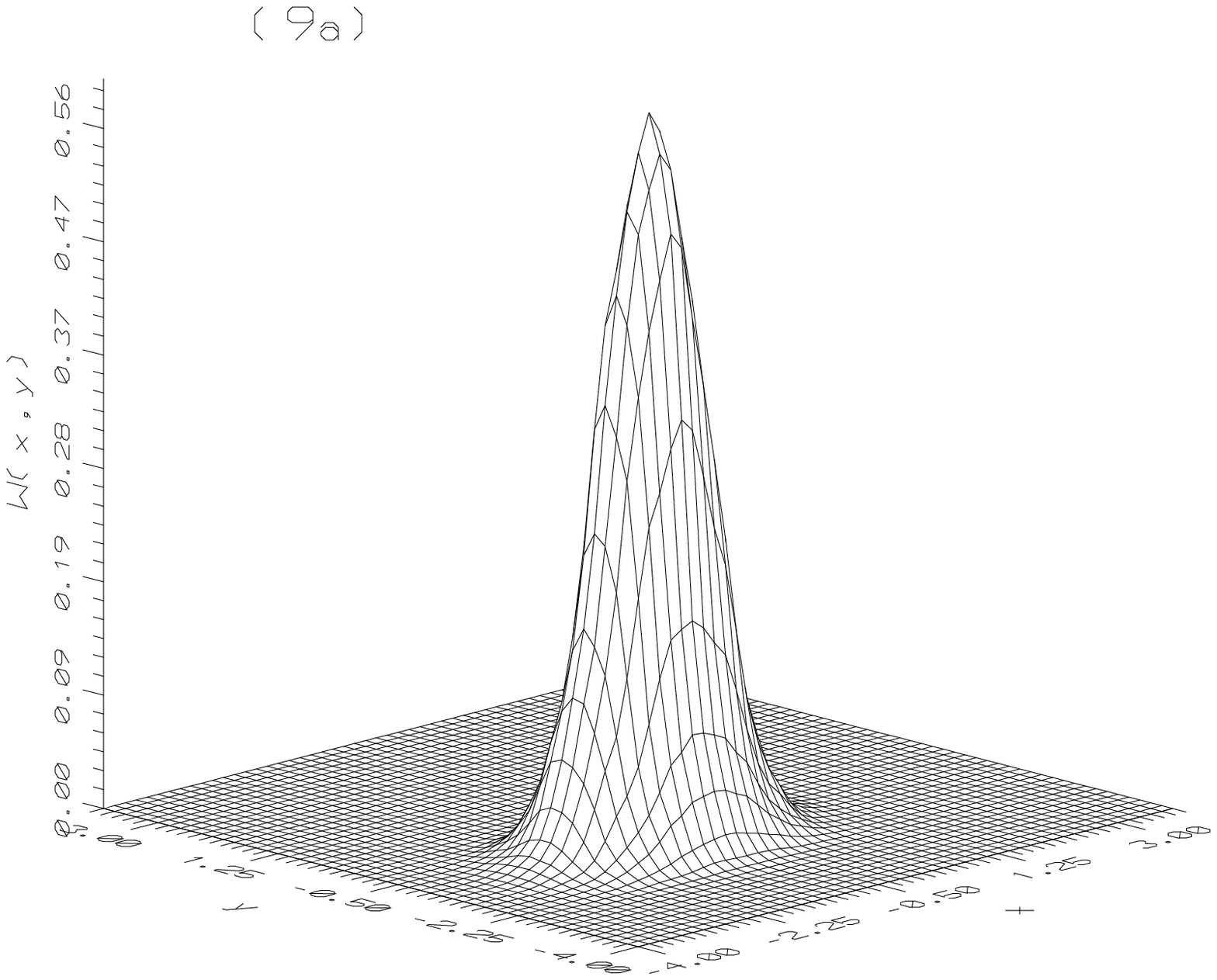}}
 \subfigure[]{\includegraphics[width=8cm]{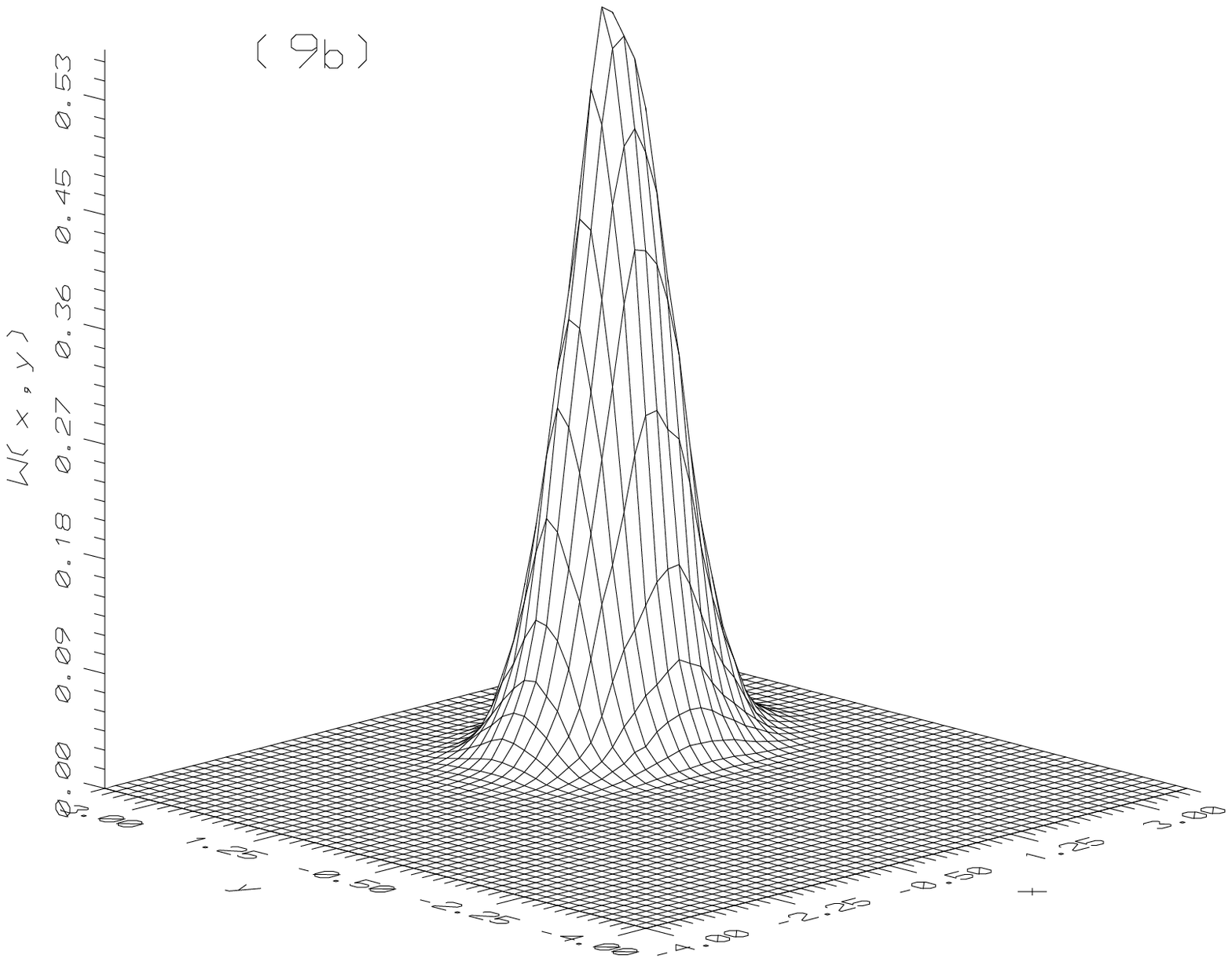}}
     \caption{
W-function for the single mode (mode 1) for different values
of time $t$ when both the modes are initially in the coherent states; $%
|\alpha_{1}|^{2}=|\alpha_{2}|^{2}=2$ and $\lambda_{k}$ are the
same as in Fig. 7: a) for $t=\pi$; b) for $t=2\pi$.   }
  \label{fig10}
\end{figure}
Furthermore, the nonclassical effect, especially squeezing of vacuum
fluctuations in the case of our
system, can be
recognized in the behaviour of $W$-function (and/or $Q$-function) in phase
space as shown in Fig. 9 for shown values of parameters.
For $t=0$, i.e. when there is no
interaction between the two modes, the W-function is identical with that
shown for a single mode representing a symmetric Gaussian bell in
phase space. As soon as the interaction switches
on ($t > 0$), we observe that the Gaussian centre is shifted and the
rotationally
symmetric function of the initial state at $t=0$ gets to be squeezed in
various phase space directions in dependence on time, as demonstrated in
Figs. 9a,b. In other words, the initial symmetric contour of the
$W$-function has been stretched as the interaction switches on, i.e. noise
ellipse characterizing squeezed light appears, which
rotates in  the phase space as the interaction time  progresses.\newline

{\bf (iii) Input thermal light} \newline
Signal beams are usually accompanied by thermal noise, so that examination
of quantum fields with thermal noise is an important problem from both
theoretical and practical points of view.
Such
thermal field can be generated by a thermal source composed of many
independent atomic radiators and consists of the superposition of waves
of
many different frequencies within some continuous range. These waves can
be
regarded as independent waves with random phases \cite{[16]}.
This field possesses uniform phase distribution (it is described
by normal distribution), exhibits thermal
statistics, i.e. $g^{(2)}(0)=2$, and its photon distribution is
the Bose-Einstein distribution.
\begin{figure}[h]%
  \includegraphics[width=8cm]{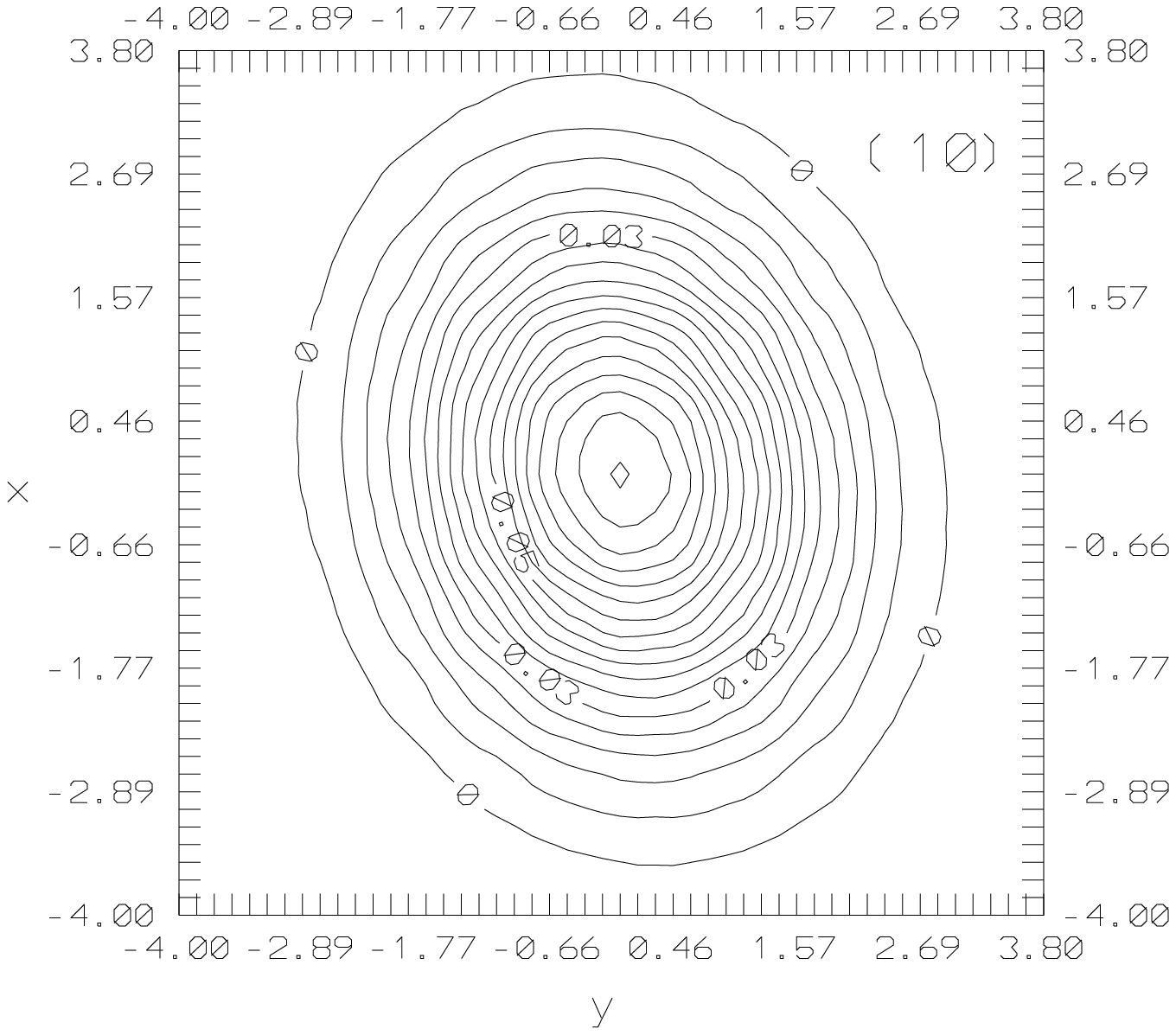}
     \caption{
The cut of the $W$-function for the single mode (mode 1) when both
the modes are initially in the thermal states with
$\bar{n}_{1}=\bar{n}_{2}=\sqrt{2},\lambda_{1}=\lambda_{2}=\lambda_{4}=0.25$
, $\lambda_{3}=1$ and $t=\frac{\pi}{2}$.  }
  \label{fig11}
\end{figure}
Here we study the
quasiprobability
functions for two modes as well as for a single mode as before, when both
the
modes are thermal. In this case the density operator takes the form

$ {\displaystyle \hat{\rho}_{T}(0)= \frac{1
}{(\bar{n}_{1}+1)(\bar{n}_{2}+1)}
\sum_{n,m=0}^{\infty} \left(
\frac{\bar{n}_{1}}{\bar{n}_{1}+1}\right)^{n}
\left( \frac{\bar{n}_{2}}{\bar{n}_{2}+1}\right)^{m} {\rm
|n\rangle_{1}|m%
\rangle_{2}}{\rm _{2}\langle m| _{1}\langle n|}, }\hfill (50) $

\noindent
where $\bar{n}_{1}\quad (\bar{n}_{2})$ is the average thermal photon number
for
mode 1 (2).
 It is clear that the thermal distribution has a diagonal
expansion
in terms of the Fock states. This diagonality causes the
electric field expectation value inside the coupler to vanish
in thermal equilibrium at all times. This, of course, is related with the
linearity of the relations (5) in terms of creation and annihilation
operators.

The two-mode $s$-parametrized characteristic function is
given
as

$ {\displaystyle C^{(2)}_{th}(\zeta_{1},\zeta_{2},s,t)=\exp \left[
-(\bar{%
n}_{1}+ \frac{1}{2}) |Z_{1}(t)|^{2}
-(\bar{n}_{2}+\frac{1}{2})|Z_{2}(t)|^{2}
+\frac{s }{2}(|\zeta_{1}|^{2} +|\zeta_{2}|^{2}) \right], }\hfill (51)
$

\noindent where

$ {\displaystyle Z_{1}(t)=\zeta_{1} K_{1}^{*}(t) -
\zeta_{1}^{*} L_{1}(t) + \zeta_{2} M_{2}^{*}(t) - \zeta_{2}^{*}
N_{2}(t), }%
\hfill (52a) $

$ {\displaystyle Z_{2}(t)=\zeta_{2} K_{2}^{*}(t) -
\zeta_{2}^{*} L_{2}(t) + \zeta_{1} M_{1}^{*}(t) - \zeta_{1}^{*}
N_{1}(t). }%
\hfill (52b) $

\noindent Therefore the two-mode $s$-parametrized
quasiprobability
function equals

$ {\displaystyle W^{(2)}_{th}(\alpha_{1},\alpha_{2},s,t)
=
\frac{1}{\pi ^{2}\sqrt{ [\bar{A}_{1}^{2}(t)-|C_{1}(t)|^{2}][\bar{A}
_{2}^{2}(t)-|C_{2}(t)|^{2}]}} }\hfill $

$\hfill$

$ {\displaystyle \qquad \times
\exp
\left\{\frac{C_{1}(t)\alpha_{1}^{2} +C_{1}^{*}(t)\alpha_{2}^{2}}
{2[\bar{A}
_{1}^{2}(t)-|C_{1}(t)|^{2}]} -
\frac{|D_{1}(t)-\alpha_{2}|^{2}}{[\bar{A}
_{2}^{2}(t)- |C_{2}(t)|^{2}]}  \right\} }\hfill $

$\hfill$

$ {\displaystyle \qquad \times \exp \left\{\frac{C_{2}^{*}(t)[D_{1}(t)-%
\alpha_{2}]^{2} +C_{2}(t)[D_{1}^{*}(t)-\alpha_{2}^{*}]^{2}} {2[\bar{A}%
_{2}^{2}(t)-|C_{2}(t)|^{2}]} \right\}, }\hfill (53) $

\noindent where we have defined

$ {\displaystyle  \bar{A}_{1}(t)=(\bar{n}_{1}+\frac{1}{2}%
)[|L_{1}(t)|^{2}+|K_{1}(t)|^{2}] +(\bar{n}_{2}+\frac{1}{2}%
)[|M_{1}(t)|^{2}+|N_{1}(t)|^{2}]-\frac{s}{2},}\hfill $

$ {\displaystyle  \bar{A}_{2}(t)=(\bar{n}_{1}+\frac{1}{2}%
)[|M_{2}(t)|^{2}+|N_{2}(t)|^{2}] +(\bar{n}_{2}+\frac{1}{2}%
)[|K_{2}(t)|^{2}+|L_{2}(t)|^{2}]- \frac{s}{2}, }\hfill $

$ {\displaystyle  C_{1}(t)=2\left[(\bar{n}_{1}+\frac{1}{2}%
)L_{1}^{*}(t)K_{1}^{*}(t)+(\bar{n}_{2}
+\frac{1}{2})M_{1}^{*}(t)N_{1}^{*}(t)
\right], }\hfill $

$ {\displaystyle  C_{2}(t)=\frac{1}{\bar{A}_{1}^{2}(t)-|C_{1}(t)|^{2}}
\left[C_{1}(t)l_{1}^{*2}(t)+ C_{1}^{*}(t)l_{2}^{2}(t)-2\bar{A}%
_{1}(t)l_{1}^{*}(t)l_{2}(t) \right] }\hfill $

$ {\displaystyle \qquad
+2\left[(\bar{n}_{2}+\frac{1}{2})L_{2}(t)K_{2}(t)+(%
\bar{n}_{1} +\frac{1}{2})M_{2}(t)N_{2}(t)\right],}\hfill $

$ {\displaystyle  D_{1}(t)=\frac{1}{\bar{A}_{1}^{2}(t)-|C_{1}(t)|^{2}}
\left\{\bar{%
A}_{1}(t)[\alpha_{1}l_{1}(t)+
\alpha_{1}^{*}l_{2}^{*}(t)]-\alpha_{1}l_{2}^{*}(t)C_{1}(t)
-\alpha_{1}^{*}l_{1}(t)C_{1}^{*}(t)\right\},}\hfill (54) $

\noindent such that $|\bar{A}_{j}(t)|>|C_{j}(t)|$.

In equation (54) we have defined $l_{1}(t)$, and $l_{2}(t)$ as
follows:

$ \displaystyle
l_{1}(t)=(\bar{n}_{1}+\bar{n}_{2}%
+1)[L_{1}^{*}(t)M_{2}^{*}(t)+K_{1}^{*}(t)N_{2}^{*}(t)], \hfill $

$ {\displaystyle
l_{2}(t)=(\bar{n}_{1}+\frac{1}{2})[L_{1}^{*}(t)N_{2}^{*}(t)+
K_{1}^{*}(t)M_{2}^{*}(t)]+(\bar{n}_{2}+\frac{1}{2}%
)[M_{1}^{*}(t)K_{2}(t)+N_{1}^{*}(t)L_{2}(t)].}\hfill (55) $

\noindent
We can see from (53) that the thermal light (classical light)
propagating through the system under consideration
can exhibit nonclassical effects,
since the $P$-function can be singular under some constrains.
Further we can see also that the nonclassical correlation between modes
is available.

For the single-mode case the $s$-parametrized characteristic
and
quasiprobability functions are

$ {\displaystyle C_{th}^{(1)}(\zeta,s,t)=\exp \left[ -|\zeta|^{2}
(J(t)-%
\frac{s}{2})+\zeta^{*2}\frac{U(t)}{2} +\zeta^{2}\frac{U^{*}(t)}{2}
\right], }%
\hfill (56) $

\noindent then the $s$-parametrized distribution function can be
written in the form

$ {\displaystyle W_{th}^{(1)}(\alpha,s,t) = \frac{1}{\pi \sqrt{
[J(t)-\frac{s%
}{2}]^{2}-|U(t)|^{2}}} }\hfill $

$\hfill$

$ {\displaystyle \qquad \times \exp
\left\{\frac{-|\alpha|^{2}[J(t)-\frac{s}{%
2}]-\frac{1}{2}[U(t)\alpha^{*2} +U^{*}(t)\alpha^{2}]}
{[J(t)-\frac{s}{2}%
]^{2}-|U(t)|^{2}} \right\}, }\hfill (57) $

\noindent where we have denoted

$ {\displaystyle U(t)=L_{1}^{*}(t)K_{1}^{*}(t)(2\bar{n}_{1}+1) +
M_{1}^{*}(t)N_{1}^{*}(t)(2\bar{n}_{2}+1), }\hfill $

$ {\displaystyle J(t)=[|L_{1}(t)|^{2}+|K_{1}(t)|^{2}]\bar{n}_{1} +
[|M_{1}(t)|^{2}+|N_{1}(t)|^{2}]\bar{n}_{2} +\frac{1}{2}%
+|L_{1}(t)|^{2}+|N_{1}(t)|^{2}, }\hfill (58) $

\noindent with $[J(t)-\frac{s}{2}]^{2}>|U(t)|^{2}$.

It is well-known for the thermal optical cavity that photons have tendency to
bunch each other, when
photon distribution is
described by the Bose-Einstein distribution (super-Poissonian statistics).
  However, as we have shown in section 3   the single mode thermal
  light can display
squeezing of thermal fluctuations under this interaction,
e.g.
 one can derive that the coupler is the source for squeezed
 thermal light. This can also be recognized in the behaviour of
 $W$-function (see Fig. 10
where the cut through the $W$-function is displayed). In this figure
one can see  the noise ellipse for squeezed thermal light with
the center at the origin.

\noindent {\bf 6. Conclusions}

In this paper we have examined the quantum statistical properties of
radiation generated and propagated in the nonlinear optical coupler
composed
of two nonlinear waveguides operating by the second subharmonic
processes,
coupled linearly by evanescent waves and nonlinearly by nondegenerate
optical parametric process. We have demonstrated regimes for generation
and
propagation of nonclassical light exhibited by squeeezing of vacuum
fluctuations and/or antibunching of photons (sub-Poissonian photon
statistics). We have also obtained quasidistribution functions
for the initial light beams which are in coherent states, Fock states
and
thermal states. Compared to earlier results for nonlinear optical
couplers
we have shown that the nonlinear coupling increases in general quantum
noise
in the device even if in some cases it can support generation of
nonclassical light.

The motivation for examination of the system under consideration
arises from
the previous investigations of the nonlinear couplers as promising
 devices to produce nonclassical light.
When coherent light is injected initially in the system, squeezed as well
as
sub-Poissonian light can be generated.
For injected number states,
 squeezed vacuum states are produced.
When thermal light initially enters the coupler,  the
coupler can operate
as a microwave Josephson-junction parametric
amplifier
\cite{yur}.
These effects have been recognized to result from the
competition between
linear and nonlinear properties of the system and are dependent
on the initial amplitudes of the input fields.
The crucial role plays here the mechanism of the energy exchange
between waveguids.

{\bf Aknowledgments}

J. P. and F. A. A. E-O. aknowledge the partial support from the Project
VS96028 and CEZ: J14/98 of Czech Ministry of Education. One of us (M.S.A.)
is greatful for the financial
support
from the project Math 1418/19 of the Research Centre, College of
Science, King
Saud University.

\end{document}